\documentclass{elsarticle}
\usepackage{graphicx} 
\usepackage{amsmath}
\usepackage{amssymb}
\usepackage{amsfonts}
\usepackage{natbib}
\usepackage{IEEEtrantools}
\usepackage{appendix}

\usepackage{color}

\biboptions{sort&compress}

\title{Updated line list for the principal isotopologue of carbon monoxide}
\author{Vladimir G. Ushakov, Emile S. Medvedev}
\address{Federal Research Center of Problems of Chemical Physics and Medicinal Chemistry (former Institute of Problems of Chemical Physics), Russian Academy of Sciences, 142432 Chernogolovka, Russian Federation}

\date{June 2025}

\begin{document}

\begin{abstract}
    The line list for the principal isotopologue of CO calculated earlier by the present authors \cite{Medvedev22preprint,Medvedev22} with the irregular dipole-moment function (DMF) is updated with use of the recent high-precision measurements in the 3-0 \cite{Bielska22,Hodges25} (Bielska \emph{et al.} 2022, Hodges \emph{et al.} 2025) and 7-0 \cite{Balashov23} (Balashov \emph{et al.} 2023) bands. The new data came in contradiction with 
    the experimental data on the 1-0 band \cite{Zou02,Devi18}. Therefore, we performed fitting several model DMFs to the modified original data set of Meshkov \emph{et al.} \cite{Meshkov22} by including the new above-referenced data 
    and by excluding the data for the 1-0 band. The updated line list is calculated with the irregular DMF. In particular, excellent agreement with recent high-level \emph{ab initio} calculations  on the 3-0 band \cite{Bielska22} is emphasized and predictions for the 1-0 and 8-0 bands are outlined. In the new update of the HITRAN database \cite{Gordon25}, new high-precision measurements in the cold and hot fundamental bands are announced. When these data are published, they will be compared with the predictions of our new line list.
\end{abstract}

\maketitle

\section{Introduction}

Recently, unprecedented high-precision intensity measurements of the 3-0 band transitions have been reported by Hodges et al. (2025) \cite{Hodges25} (\textbf{Hodges25}), who have also incorporated and analyzed the earlier data of Bielska et al. (2022) \cite{Bielska22} (\textbf{Bielska22}), and the 7-0 band transitions have been measured for the first time by Balashov et al. (2023) \cite{Balashov23} (\textbf{Balashov23}). These new data pose two questions: 1) how our previously published calculated intensities, Meshkov {et al}. \cite{Meshkov22}  (\textbf{Meshkov22}), Medvedev and Ushakov \cite{Medvedev22} (\textbf{Medvedev\&Ushakov22}), conform with these new data? 2) can the line list be improved significantly by making new fittings with use of these new data?

\section{Preliminary remarks}

We start with a discussion of the 1-0 band because of its special role in the new fittings. Our published line lists \cite{Medvedev22preprint,Medvedev22,Meshkov22} were based on the fittings that used the data of Zou and Varanasi (2002) for a 99.97\% isotope-enriched sample \cite{Zou02} (\textbf{Zou02}) and Devi \emph{et al.} (2018) for a natural sample \cite{Devi18} (\textbf{Devi18}). Both these measurements gave the intensities (with the \textbf{Zou02} data recalculated from abundance of 0.9997 to the natural abundance of 0.986544 as specified in HITRAN) \emph{weaker} by about 2\% than in HITRAN2016. 
Therefore, in HITRAN2020, the 1-0 band intensities were \emph{decreased} by a factor of 1.02
. However, the new measurements announced in Ref. \cite{Gordon25} better agree with HITRAN2016. In this paper, using the new data, we will try to fit the \textbf{Medvedev\&Ushakov22} irregular dipole-moment function both with and without the above-mentioned previous data on the 1-0 band.

Before description of the fittings with use of the new data, a notice about the fitting procedure is in order. In \textbf{Meshkov22} and \textbf{Medvedev\&Ushakov22}, a total of 508 experimental intensity values in various bands out of about 1500 measured lines were selected for the fitting of the regular and irregular DMFs, respectively. The experimental intensities were converted into the transition dipole moments (TDMs), which were actually fitted. The \emph{ab initio} values of the DMF were also used in the fitting with the error bars as defined in \textbf{Meshkov22}. 

One issue with the new fittings was that some data could not be fitted simultaneously because of probable systematic errors that conflicted with each other. The other one concerned with the uncertainties of the \emph{ab initio} data, which are usually unknown.
Increasing the uncertainties can diminish the weight of the theoretical data as compared to the experimental ones. For the calculations of the high-overtone intensities, very precise \emph{ab initio} data are required. However, due to the wealth of the experimental data, we may not use the \emph{ab initio} data in the fitting. The actual choice of the fitting strategy is discussed in the next section.

\section{Fittings}\label{Sec:fit}

Our fitting method uses several model DMFs with different analytic properties in order to satisfy the obvious requirement that the calculated intensities were more or less independent of the model used. 

We will fit three model DMFs, $d(r)$: the regular \cite{Meshkov22} and irregular \cite{Medvedev22preprint,Medvedev22} DMFs mentioned above plus a rational one \cite{Medvedev22} having five pairs of complex-conjugated poles in the complex plane. All of them satisfy the following limiting conditions: $d(r)\propto r^3$ at $r\rightarrow0$ and $d(r)\approx d_4r^{-4}$ at $r\rightarrow\infty$ with $d_4$ fixed at $-5.1$ DÅ$^4$, if not specified otherwise.

\begin{figure}[htbp]
    \centering
    \includegraphics[scale=0.25]{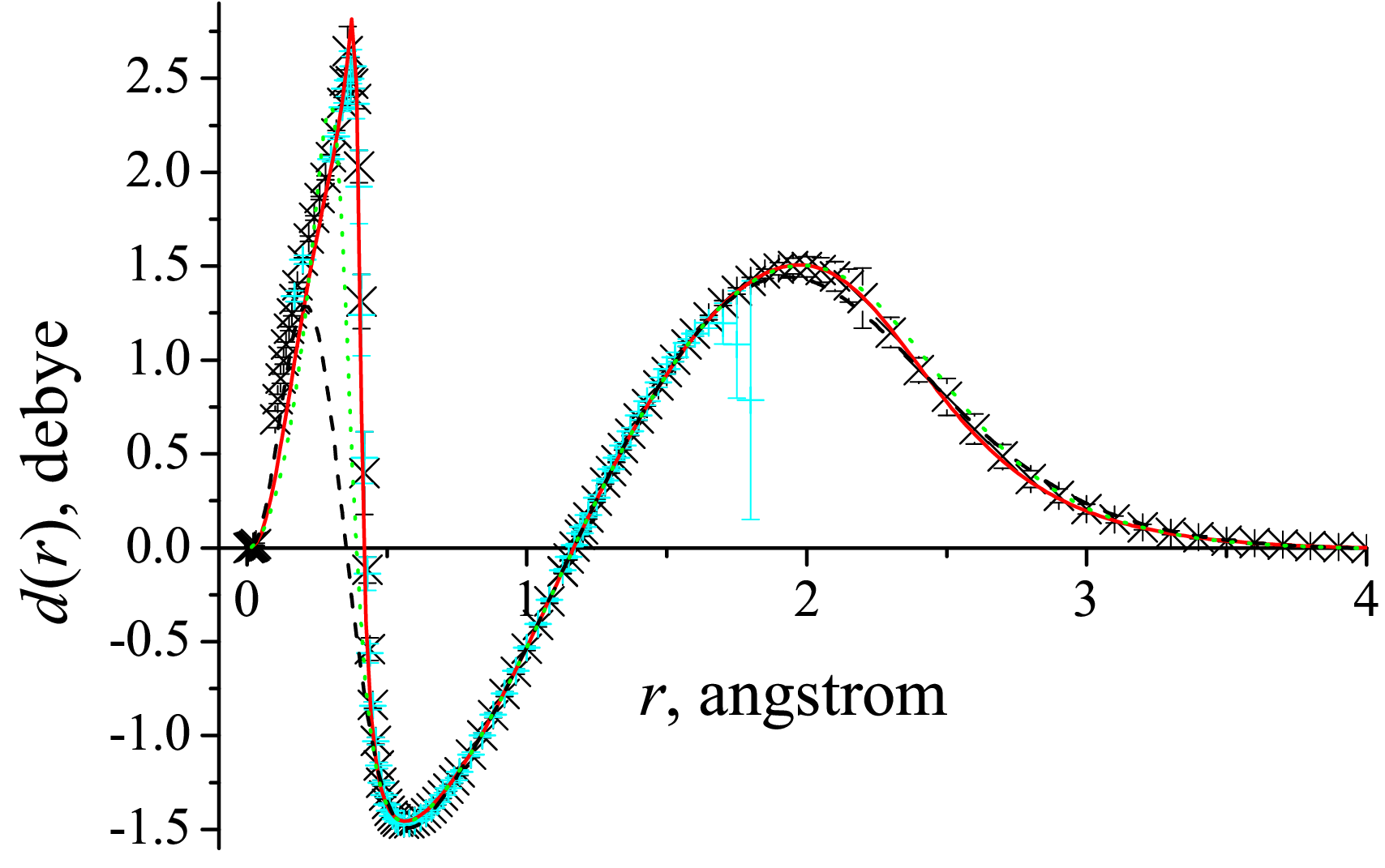}
    \caption{The DMFs obtained in Fit 0. Solid line, the irregular DMF with $d_4$ fixed; dashed line, the regular DMF with $d_4$ fixed; dotted line, the rational DMF with $d_4$ not fixed (see text). Crosses and pluses, the \emph{ab initio} (ACPF and CCSD(T), respectively) data of \textbf{Meshkov22} with the original error bars.}
    \label{fig:DMFs_Fit0}
\end{figure}

In Fit 0, we try to accommodate the new data within the model of \textbf{Medvedev \&Ushakov22}. Thus, the intensity data by \textbf{Balashov23} for the 7-0 band and \textbf{Hodges25} for the 3-0 band ($\Bar{S}_m$ from Table 8.2) recalculated to the TDMs are added to the original data set used in \textbf{Meshkov22}. The fitting results are shown in Figs. \ref{fig:DMFs_Fit0}-4.

Figure \ref{fig:DMFs_Fit0} shows the three DMFs with fitted parameters. In fitting the rational DMF, parameter $d_4$ was variable to obtain $d_4=-3.9$ D\AA$^4$. 

As expected, all DMFs closely coincide with each other within the spectroscopically accessible region, $\approx 0.8$-1.9 \AA, but diverge outside it, especially at short $r$. The rational function was fitted with floating $d_4$, yet fitting with the fixed $d_4$ little affected both the DMF shape and the intensities. The regular DMF poorly reproduces the \emph{ab initio} data at $r<0.5$ \AA{} despite the fact that these data were included in the fitting. The overall quality of the fit is analyzed below in the comparison of the calculated and observed intensities.

Such comparisons for the 3-0 and 7-0 bands are shown in Figs. \ref{fig:Band_3-0,Fit0} and \ref{fig:Band_7-0,Fit0}, respectively. It is seen that the 3-0 band is poorly reproduced by the model DMFs. The 7-0 band is reproduced by all DMFs within the experimental errors but there are systematic displacements for the irregular and rational DMFs. There are two possible reasons for such shortcomings: the errors in the \emph{ab initio} DMF and inconsistencies in the experimental data for various bands due to systematic errors. In particular, such systematic errors are clearly seen in the \textbf{Devi18} data for the 1-0 band shown in Fig. 4. Thus, we consider Fit 0 unsatisfactory.

\begin{figure}[htbp]
    \centering
    \includegraphics[scale=0.25]{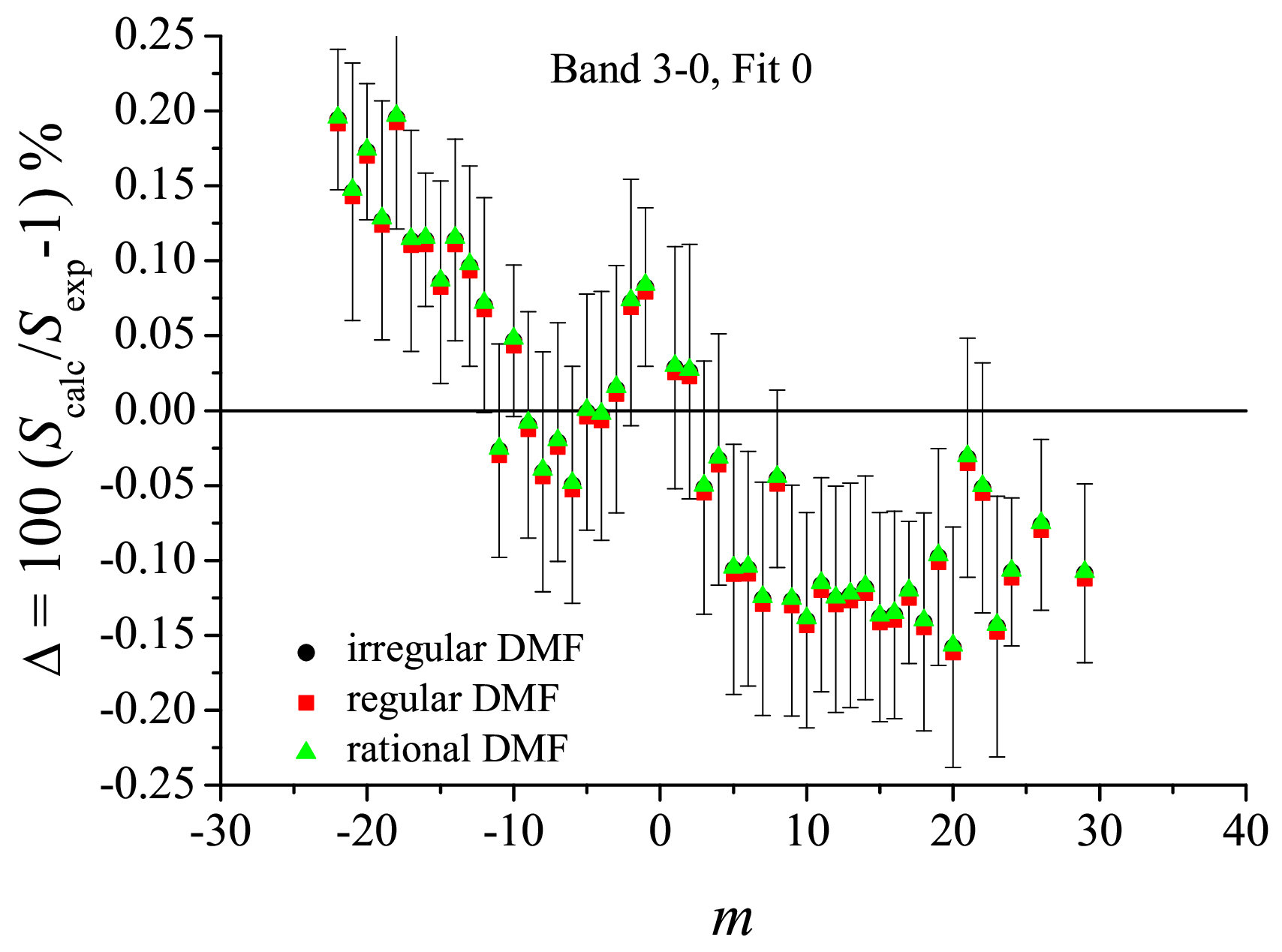}
    \caption{Comparison of the calculated (Fit 0) and measured (\textbf{Hodges25} \cite{Hodges25}) intensities in the 3-0 band. Here and below, the error bars show the experimental uncertainties.}
    \label{fig:Band_3-0,Fit0}
\end{figure}

\begin{figure}[htbp]
    \centering
    \includegraphics[scale=0.25]{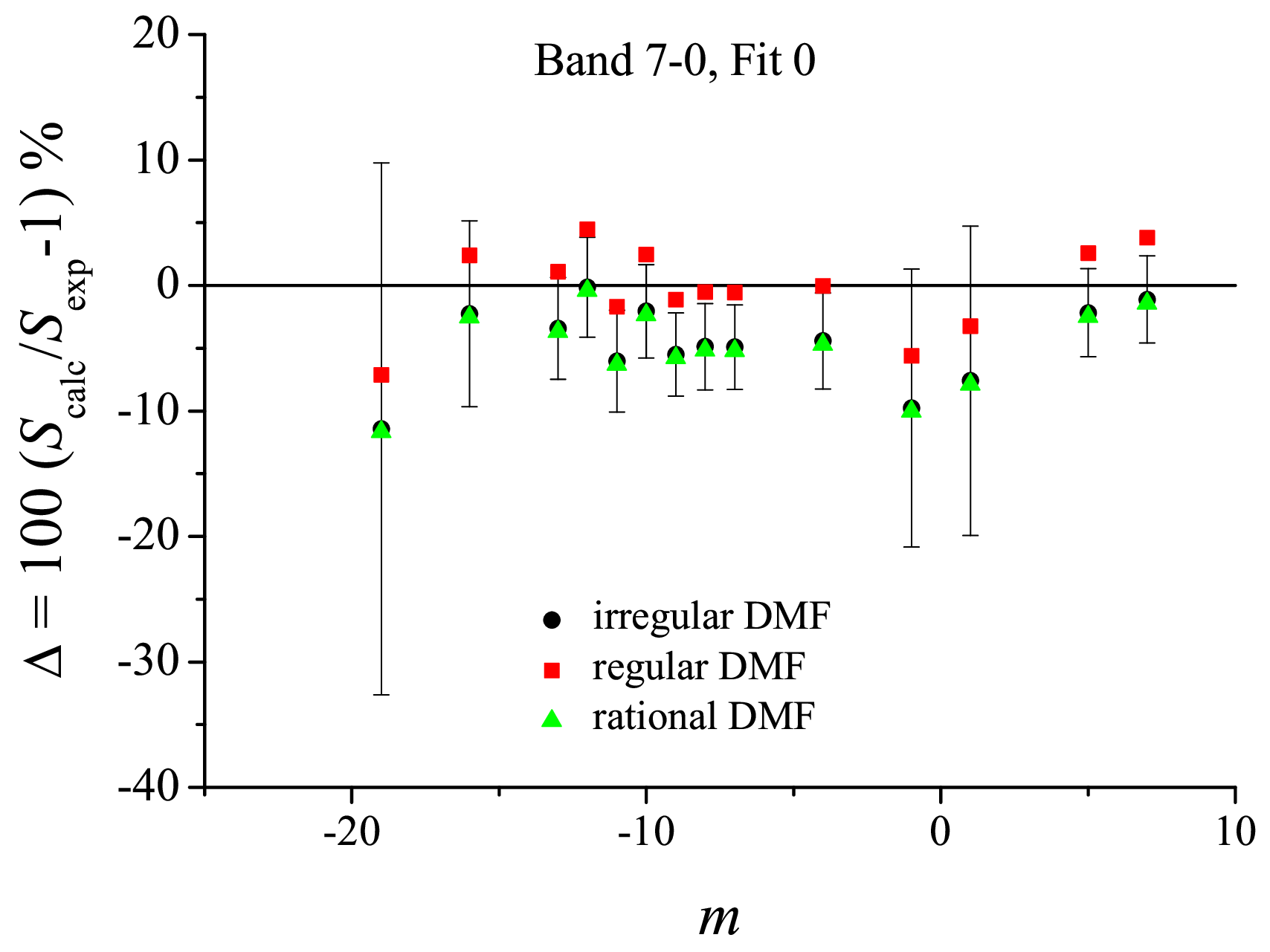}
    \caption{Comparison of the calculated (Fit 0) and measured (\textbf{Balashov23} \cite{Balashov23}) intensities in the 7-0 band.}
    \label{fig:Band_7-0,Fit0}
\end{figure}

\begin{figure}[htbp]
    \centering
    \includegraphics[scale=0.25]{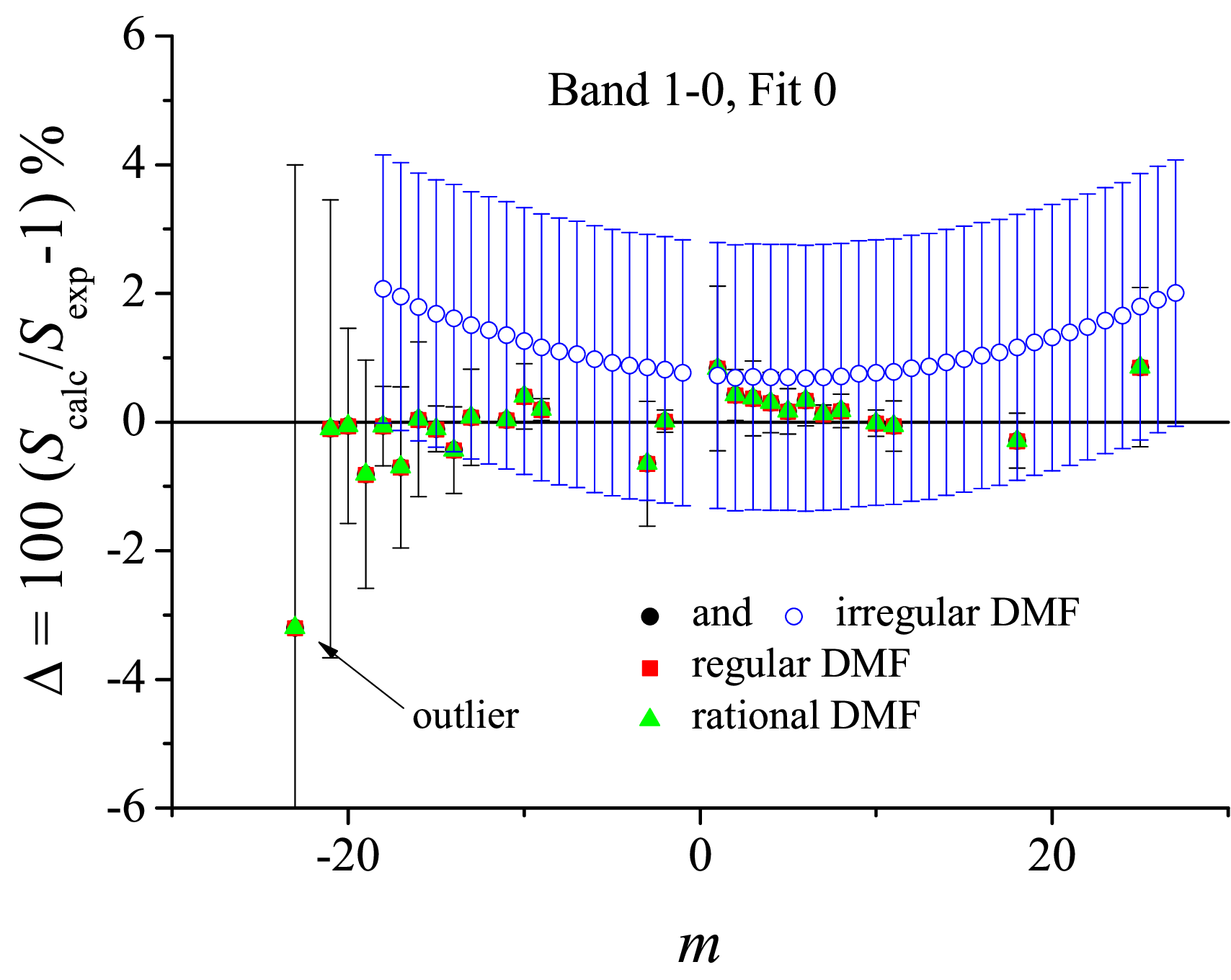}
    \caption{Comparison of the calculated (Fit 0) intensities in the 1-0 band with the experiment of \textbf{Zou02} (filled symbols) and \textbf{Devi18} (open symbols). The results for all DMFs are identical.}
    \label{fig:Band_1-0,Fit0}
\end{figure}

In Fit 1, we investigate how the \emph{ab initio} data affect the model DMF behavior and the calculated intensities. In principle, we can use solely the experimental data for the fitting because they are sufficiently rich. Such a fitting results in all three DMFs giving similar intensities. However, the regular DMF strongly declines from the \emph{ab initio} points and turns down at $r<0.5$ \AA. In order to avoid such a non-physical behavior, we restore the fitting to the \emph{ab initio} DMF of \textbf{Meshkov22} but the original uncertainties are strongly increased: $\sigma$ = 200 $\sigma_\textrm{orig}$ at $r < 0.5$ \AA{} and $\sigma$ = max(0.3 D,$10\sigma_{\textrm{orig}}$) at $r > 0.5$ \AA.

\begin{figure}[htbp]
    \centering
    \includegraphics[scale=0.25]{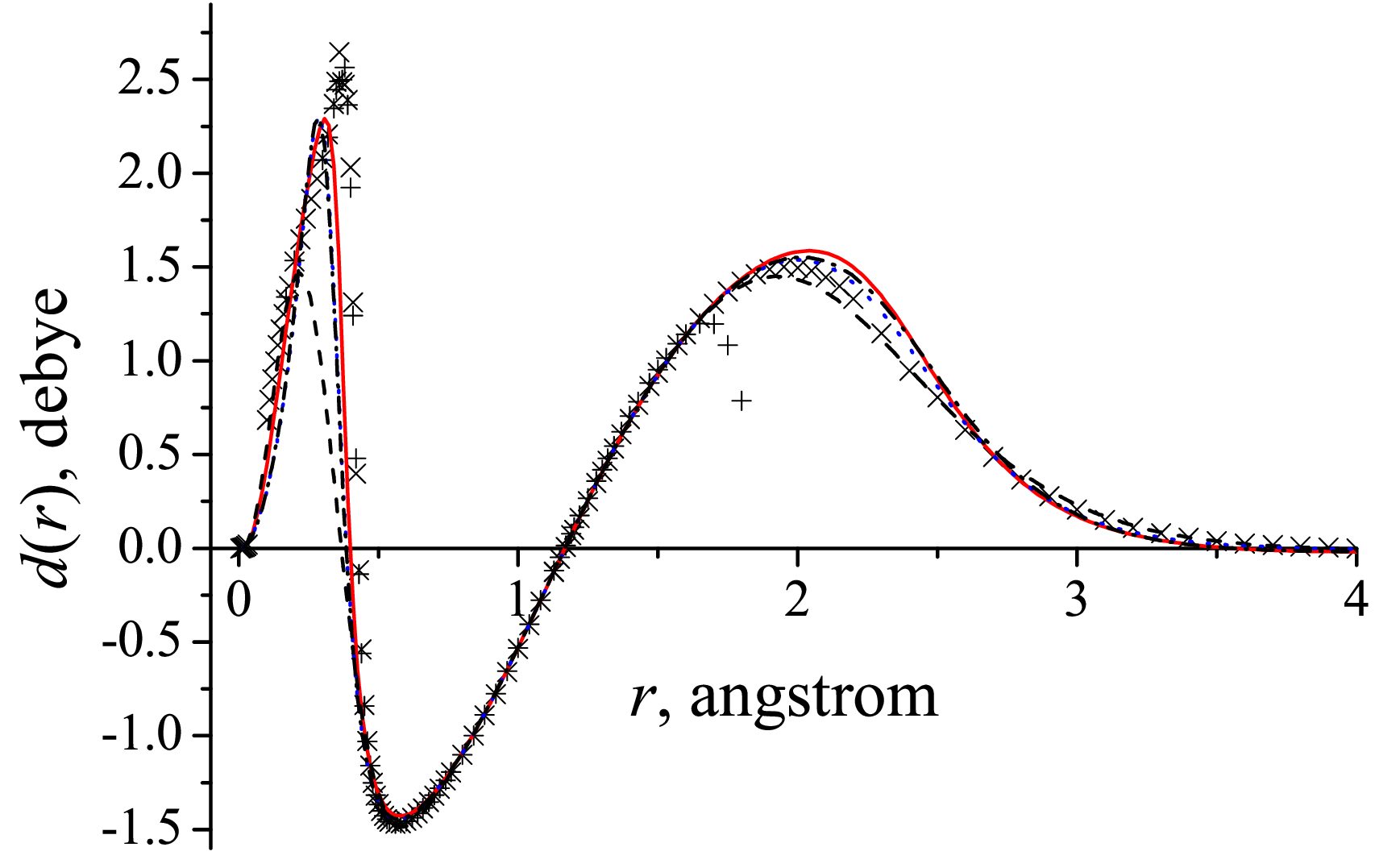}
    \caption{The DMFs obtained in Fit 1. Crosses and pluses show the \emph{ab initio} DMFs calculated by ACPF and CCSD(T) methods, respectively. The increased uncertainties of the \emph{ab initio} data are not shown. Full and dashed lines present the irregular and regular DMFs, respectively. The rational DMF is present in two forms, the one with a fixed $d_4$ (dots) and the other with variable $d_4$ (dash-dot); both the shape of this DMF and the intensity results are nearly the same. }
    \label{fig:DMFs_Fit1}
\end{figure}

Three fitted DMFs are shown in Fig. \ref{fig:DMFs_Fit1} along with the \emph{ab initio} one. All functions, except for the regular one at short $r$, excellently reproduce the \emph{ab initio} data. Comparison of the intensities in the 3-0 band given in Fig. \ref{fig:Band_3-0,Fit1} shows that the systematic errors increasing with $|m|$ still manifest as in Fig. \ref{fig:Band_3-0,Fit0}. The 7-0 band is reproduced satisfactorily in Fig. \ref{fig:Band_7-0,Fit1}, \emph{i.e.} the systematic shifts down seen in Fig. \ref{fig:Band_7-0,Fit0} is now removed.

\begin{figure}[htbp]
    \centering
    \includegraphics[scale=0.25]{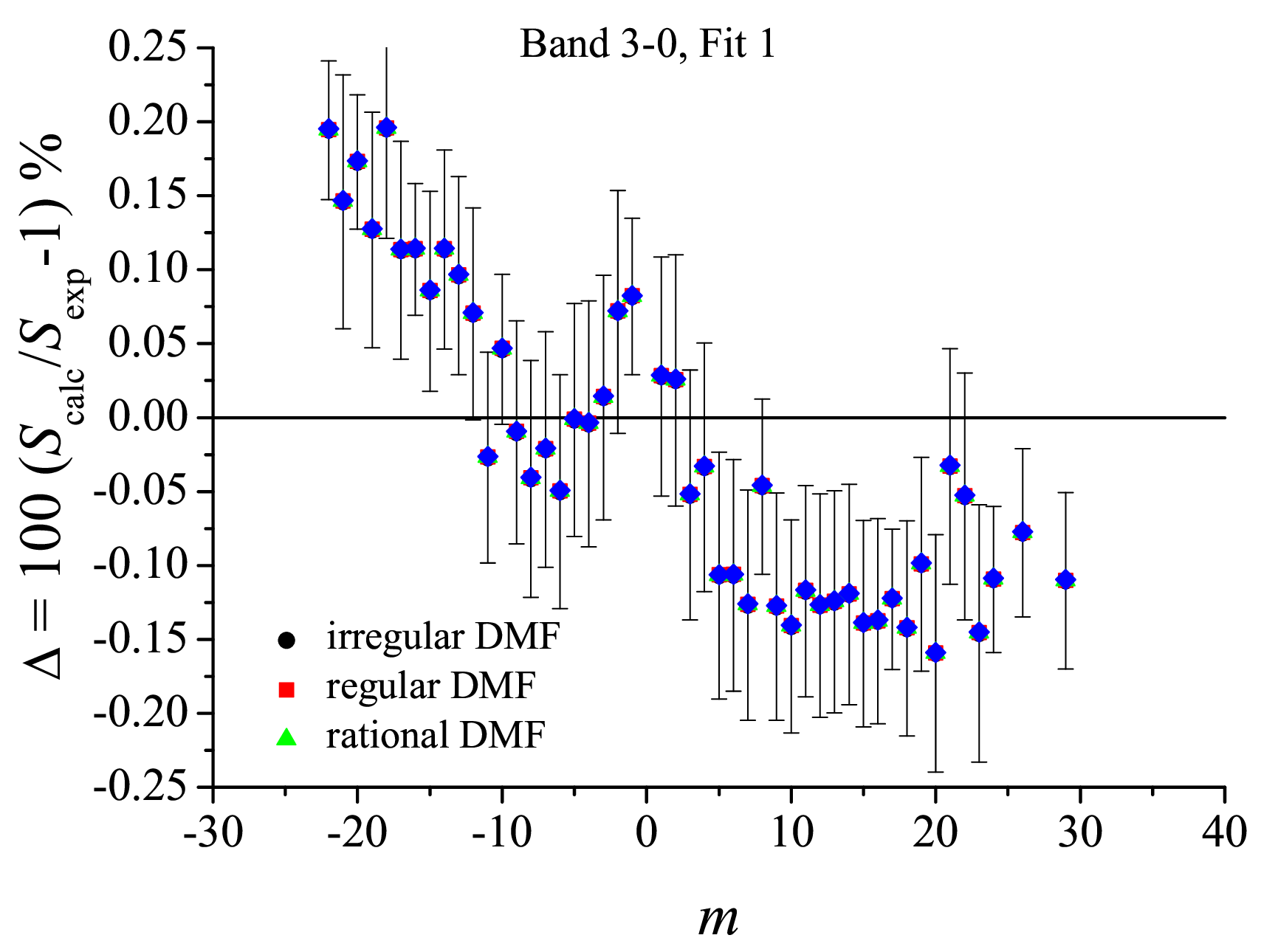}
    \caption{Comparison of the calculated (Fit 1) and measured intensities in the 3-0 band. The results for all DMFs are identical.}
    \label{fig:Band_3-0,Fit1}
\end{figure}

\begin{figure}[htbp]
    \centering
    \includegraphics[scale=0.25]{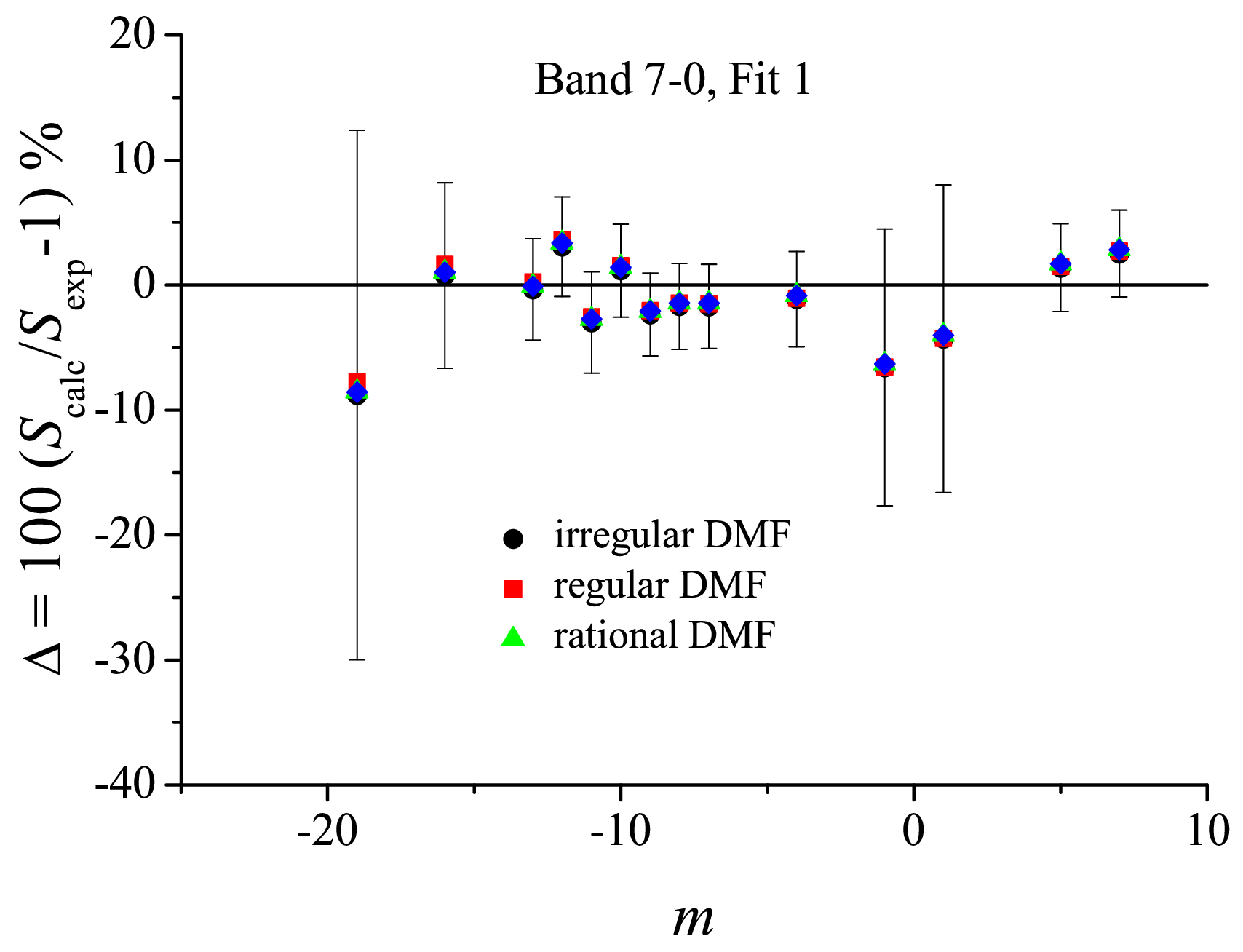}
    \caption{Comparison of the calculated (Fit 1) and measured intensities in the 7-0 band. The results for all DMFs are nearly identical.}
    \label{fig:Band_7-0,Fit1}
\end{figure}

Thus, we see that excluding the \emph{ab initio} data from the fitting or radical decreasing their weight as has been done in Fit 1 affects the high-overtone intensities, bringing the calculated 7-0 band in agreement with experiment whereas the low-overtone intensities remain unaffected. The latter also manifests for the 1-0 band intensities as shown in Fig. \ref{fig:Band_1-0,Fit1} where the same systematic shift of the \textbf{Devi18} data is present, whereas the \textbf{Zou02} data are reproduced well, all the same as in Fig. \ref{fig:Band_1-0,Fit0}.

Thus, we established in Fit 1 that the \emph{ab initio} data with the increased uncertainties must be included in the fit in order to ``repair" the calculated 7-0 band intensities and provide for the physical behavior of the regular DMF. Yet, the problems with the 3-0 and 1-0 bands remain.

\begin{figure}[htbp]
    \centering
    \includegraphics[scale=0.25]{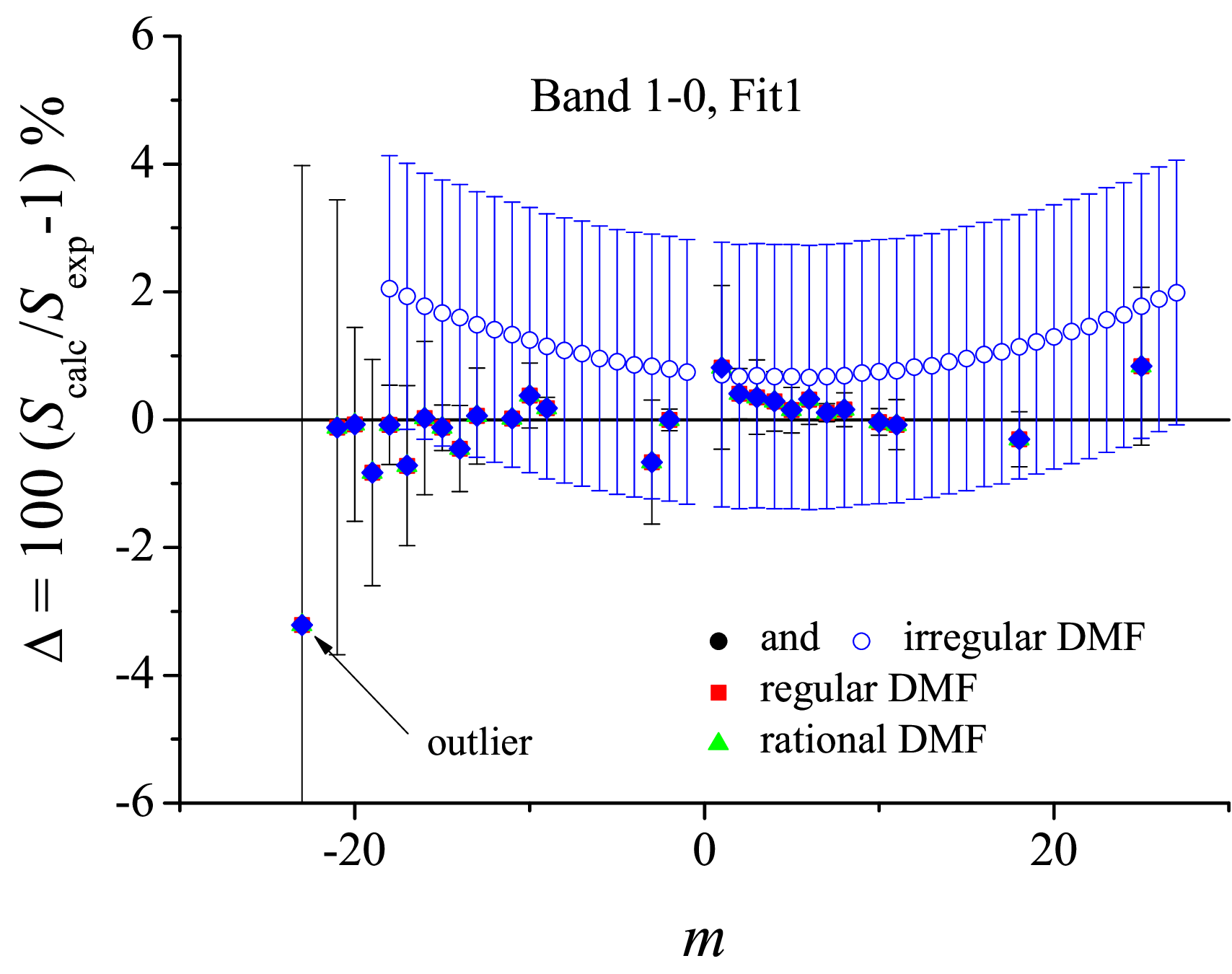}
    \caption{Comparison of the calculated (Fit 1) intensities in the 1-0 band with the experiment of \textbf{Zou02} (filled symbols) and \textbf{Devi18} (open symbols). The results for all DMFs are identical.}
    \label{fig:Band_1-0,Fit1}
\end{figure}

If we remove the \textbf{Devi18} data, the problem with the 3-0 band is not resolved whereas all other bands are reproduced within the experimental errors. There is no direct indication on which data are in contradiction with the 3-0 band. Therefore, we tried them one by one and found that these were the data of \textbf{Zou02}.

\begin{figure}[htbp]
    \centering
    \includegraphics[scale=0.25]{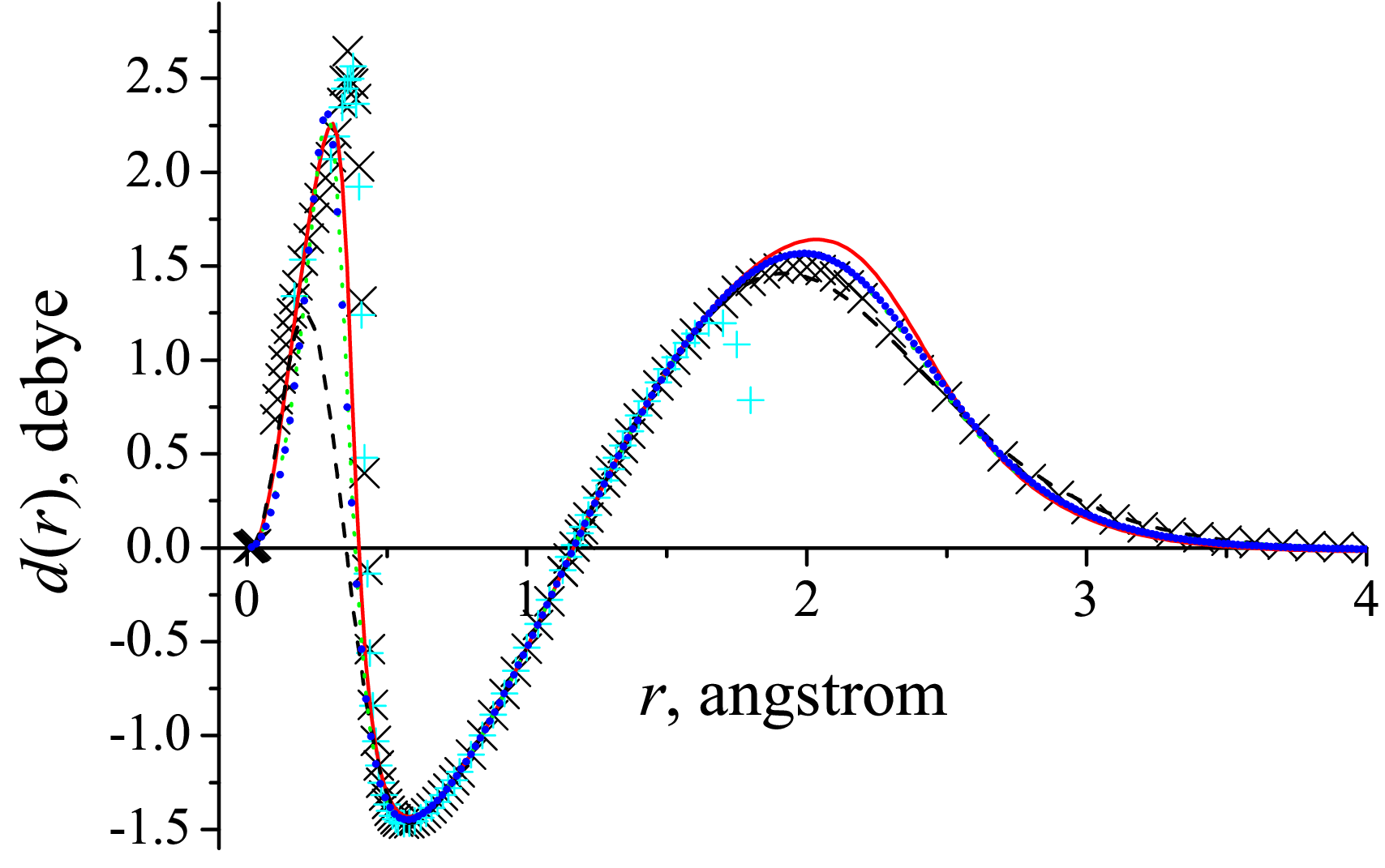}
    \caption{The DMFs obtained in Fit 2. Notations are the same as in Fig. \ref{fig:DMFs_Fit1}. The rational DMF with floating $d_4$ is not shown because it will not be used anymore.}
    \label{fig:DMFs_Fit2a}
\end{figure}

In the final Fit 2, we remove all data on the 1-0 band. Thus, the final data base (see Supplementary material) includes the following: 1) the data base of \textbf{Meshkov22} without the 1-0 band and the uncertainties of the \emph{ab initio} DMF values are increased as specified above; 2) the new data of \textbf{Hodges25} and \textbf{Balashov23} on the 3-0 and 7-0 bands. The results are shown in Figs. \ref{fig:DMFs_Fit2a}-\ref{fig:Band_1-0,Fit2} and discussed in the next section.

\begin{figure}[htbp]
    \centering
    \includegraphics[scale=0.25]{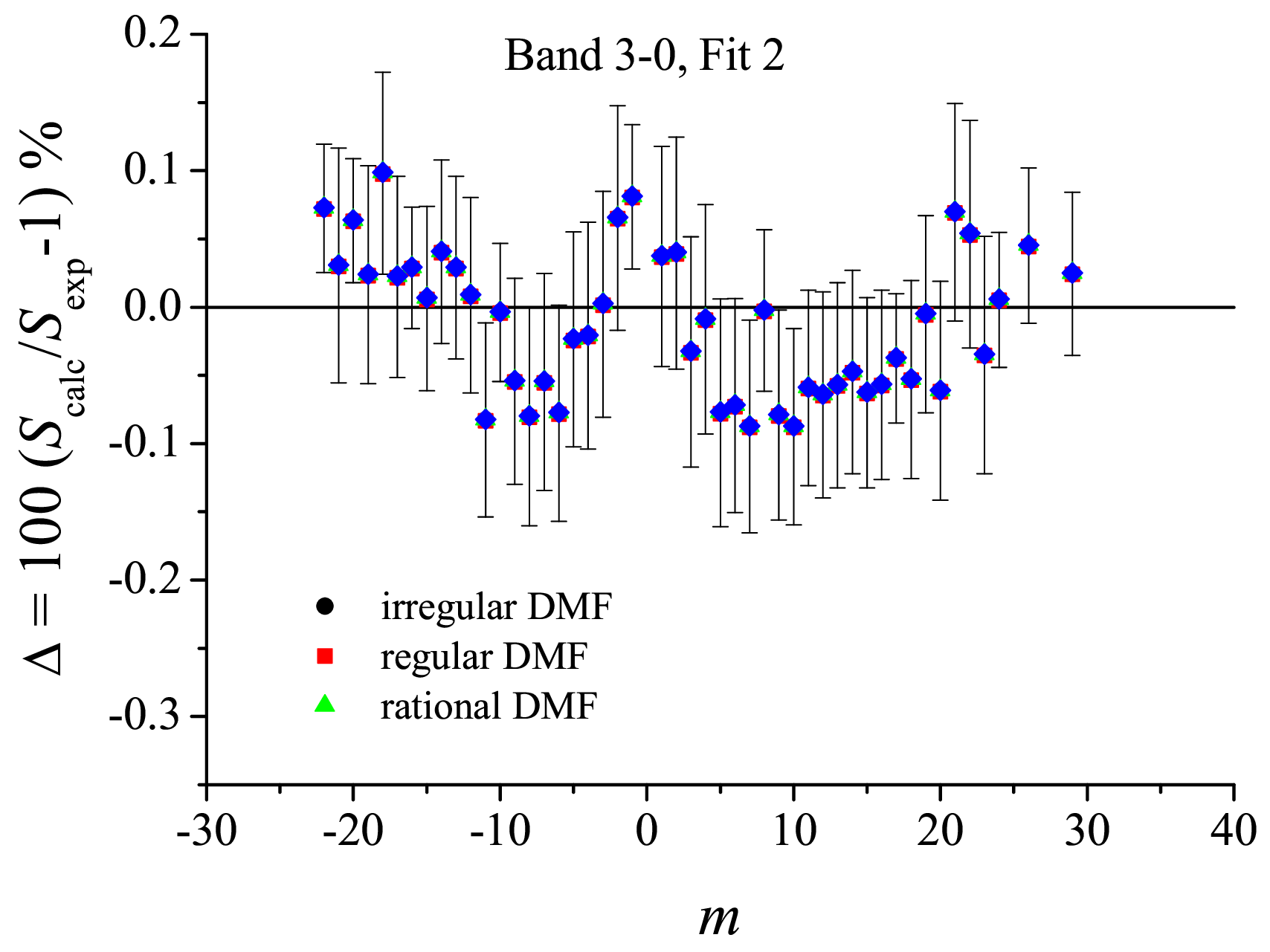}
    \caption{Comparison of the calculated (Fit 2) and measured (\textbf{Hodges25}) intensities in the 3-0 band. The results for all DMFs are identical and perfectly describe the experiment.}
    \label{fig:Band_3-0,Fit2}
\end{figure}

\begin{figure}[htbp]
    \centering
    \includegraphics[scale=0.25]{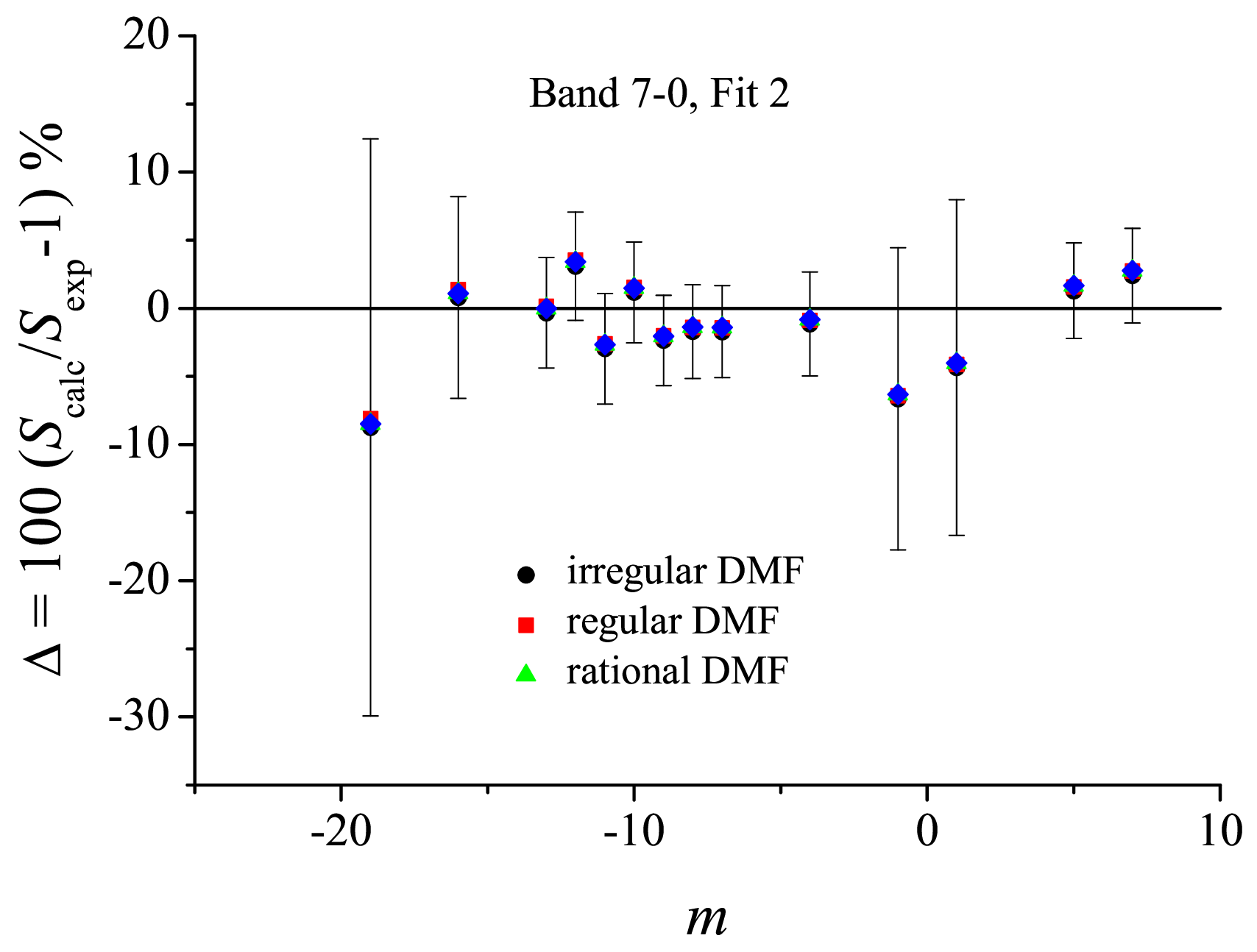}
    \caption{Comparison of the calculated (Fit 2) and measured (\textbf{Balashov23}) intensities in the 7-0 band. The results for all DMFs are identical and perfectly describe the experiment.}
    \label{fig:Band_7-0,Fit2}
\end{figure}

\begin{figure}[htbp]
    \centering
    \includegraphics[scale=0.25]{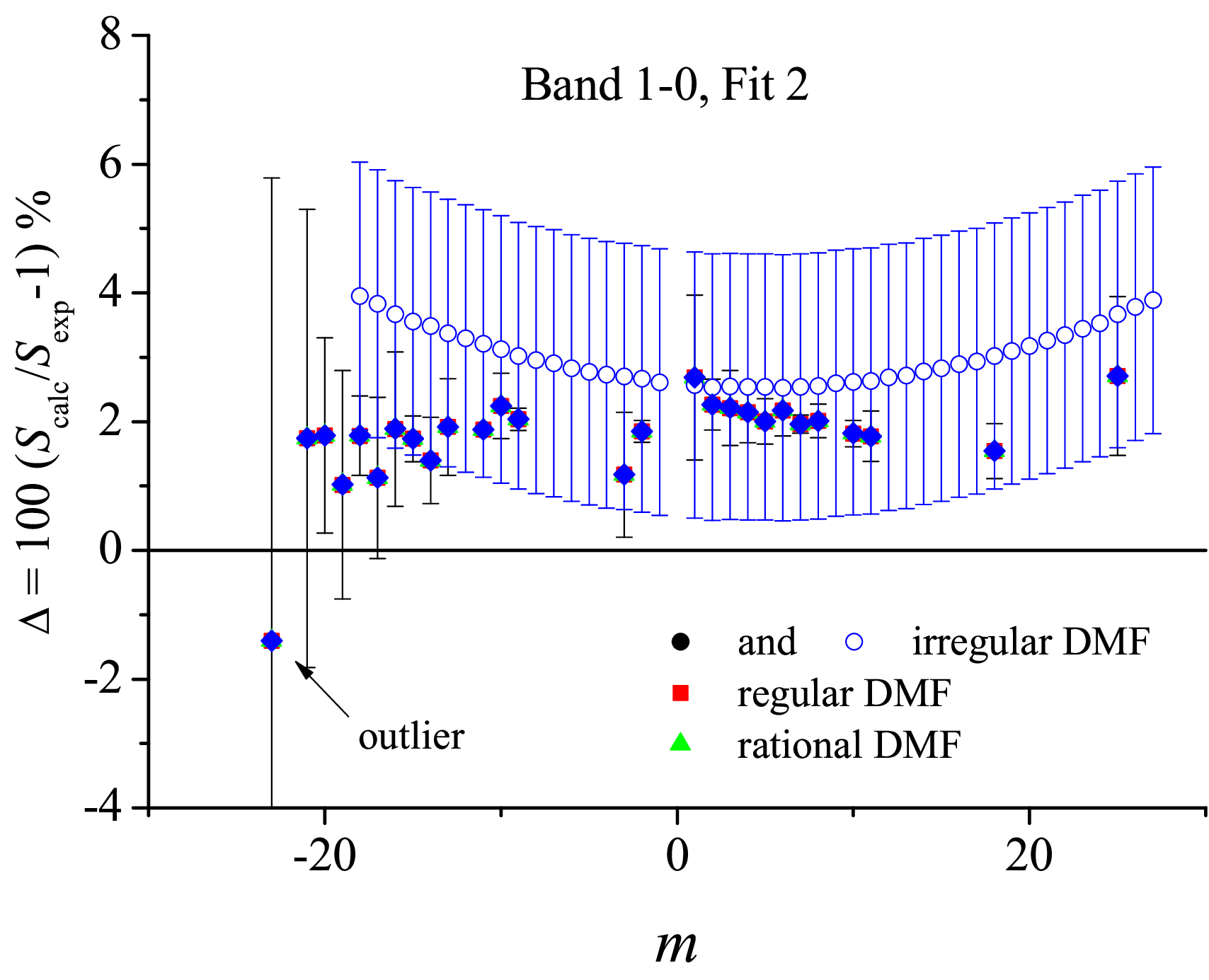}
    \caption{The 1-0 band predicted by Fit 2. The results for all DMFs are identical. Comparison of the calculated intensities in the 1-0 band with the experiments of \textbf{Zou02} (filled symbols) and \textbf{Devi18} (open symbols) is given. The experimental data show systematic errors.}
    \label{fig:Band_1-0,Fit2}
\end{figure}

\section{Analysis of the fitted functions}

The three fitted model DMFs considered in Sec. \ref{Sec:fit} are shown in Fig. \ref{fig:DMFs_Fit2a}. Qualitatively, they are the same as in Fig. \ref{fig:DMFs_Fit1}. 
All of them provide equally well description of the experimental data as seen in Figs. \ref{fig:Band_3-0,Fit2} and \ref{fig:Band_7-0,Fit2} for the 3-0 and 7-0 bands, respectively. As for the 1-0 band shown in Fig. \ref{fig:Band_1-0,Fit2}, all DMFs give similar predictions but the agreement with experiment is worse because of the systematic experimental errors.

For the irregular DMF, the quality of Fit 2 is demonstrated by Table \ref{tab:Fit_quality} where the average deviations of the calculated and observed intensities for the particular bands, $\left<\Delta\right>$, are compared with the average experimental uncertainties, $\left<\sigma\right>$. 
In the last column, the average ratios of the difference between the calculated and observed intensities to the experimental uncertainties are shown. For the majority of the bands, they are less than unity. The most of the bands, except for the 1-0 one, are described within the experimental uncertainties.

\begin{figure}[htbp]
    \centering
    \includegraphics[scale=0.25]{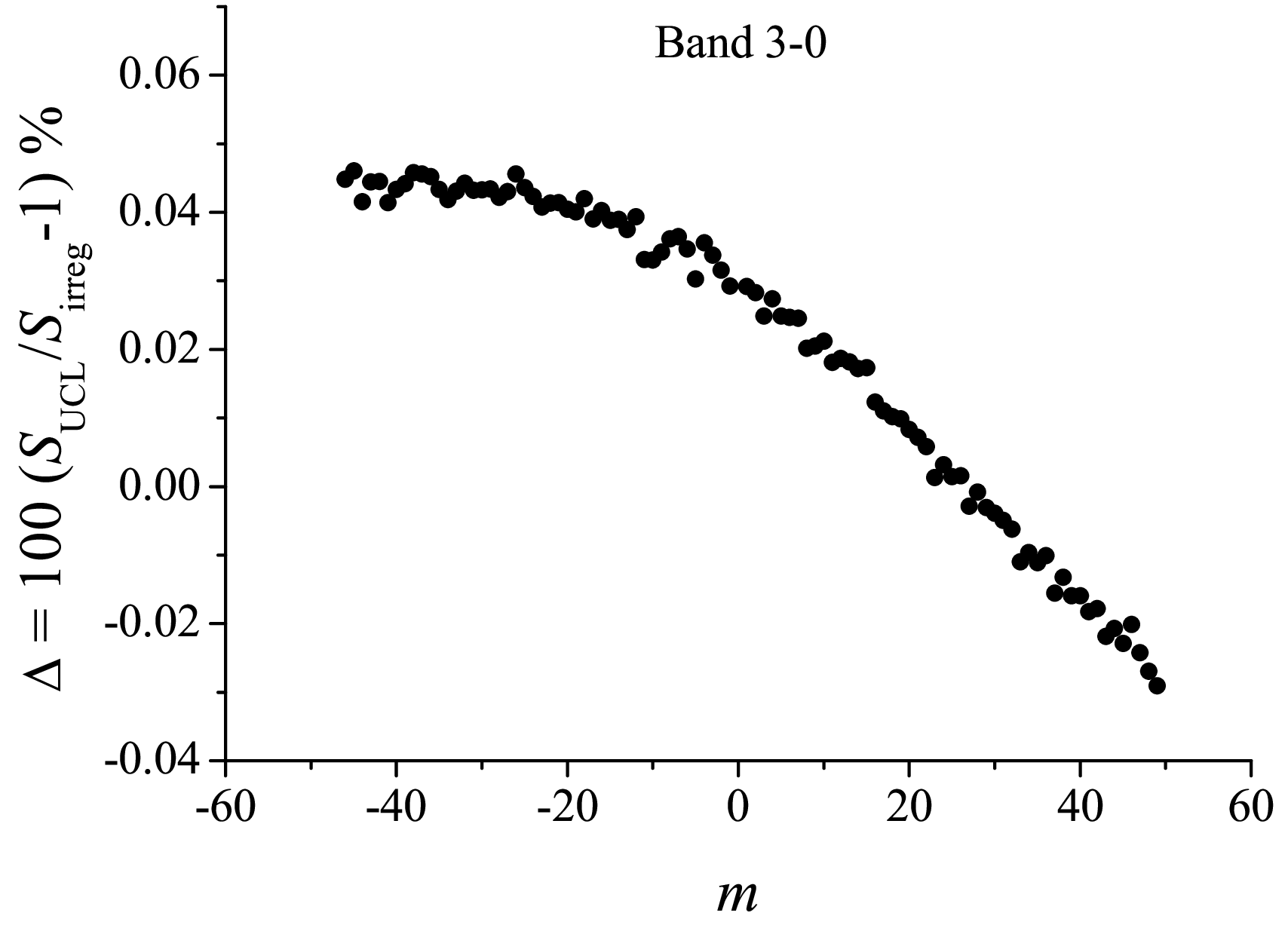}
    \caption{Difference between the 3-0 band intensities calculated in Ref. \cite{Bielska22}, $S_\textrm{UCL}$, and in the present work with the irregular DMF.}
    \label{fig:Band3-0_UCL_irreg}
\end{figure}

Figure \ref{fig:Band3-0_UCL_irreg} shows the comparison with the \emph{ab initio} data of Ref. \cite{Bielska22} for the 3-0 band. The agreement is within 0.05\% all over the band up to $J=50$. It is remarkable that two fully independent theoretical methods give the difference in the calculated intensities even less than the experimental uncertainty of 0.1\%.

However, if we want a DMF to have the predictive power, we need to impose some additional restrictions on the model functions.

\begin{table}[htbp]
    \centering
    \caption{Average values of the experimental uncertainties in the intensities, $\left<\sigma\right>$, observed-calculated differences, $\left<\Delta\right>$, and their ratios for the particular bands for the irregular DMF}
    \vspace{7pt}
    \begin{tabular}{crcrll}
  Band  & No of lines & $\left<\sigma\right>$,\% & $\left<\Delta\right>$,\% & $\left<|\Delta|/\sigma\right>$  &  Reference \\
  \hline
   0-0  &    8   &   1.18  &    0.03   &    0.25& \textbf{Aenchbacher10} \cite{Aenchbacher10} \\ 
   0-0  &   12   &   3.89  &   -1.19   &    0.43 & \textbf{Birk96}  \cite{Birk96}     \\
   0-0  &   14   &   7.44 &    -0.74&       0.21  &     \textbf{Birk96} \cite{Birk96}      \\
   1-0   &  45    &  2.07  &    3.00 &      1.45   &    \textbf{Devi18}   \cite{Devi18}$^a$     \\
   1-0    & 27     & 1.00   &   1.75  &     6.50$^b$    &   \textbf{Zou02}   \cite{Zou02}$^a$      \\
   2-0  &   57   &   0.21    &  0.02   &    0.36     &  \textbf{Devi12}  \cite{Devi12}      \\
   2-0   &  27    &  0.52     &-0.15    &   0.42      & \textbf{Loos17} \cite{Loos17}       \\
   3-0    & 52     & 0.12 &    -0.08     &  1.17     &\textbf{Bielska22} \cite{Bielska22}$^a$    \\
   3-0  &   40   &   3.01  &   -0.43      & 0.51&       \textbf{Borkov20a}  \cite{Borkov20a}    \\
   3-0   &   1    &  0.07   &   0.00       &0.04 &      \textbf{Cygan19}  \cite{Cygan19}     \\
   3-0    & 48     & 0.07    & -0.01&       0.66  &     \textbf{Hodges25}  \cite{Hodges25}    \\
   3-0  &    3   &   1.07     & 0.51 &      0.48   &    \textbf{Reed17}  \cite{Reed17}      \\
   3-0   &  45    &  2.00 &     0.77  &     0.42    &   \textbf{Sung04}    \cite{Sung04}   \\
   4-0    & 52     & 3.29  &   -0.85   &    0.47     &  \textbf{Borkov20b}  \cite{Borkov20b}    \\
   4-0  &    8 &     2.00   &  -0.03    &   0.25      & \textbf{Camparque15} \cite{Campargue15}  \\
   4-0   &  47  &    1.00    &  0.23     &  0.27       &\textbf{Bordet20} \cite{Bordet20}  \\
   4-0    & 31   &   2.00     &-0.17      & 0.40&       \textbf{Li18}  \cite{Li18}      \\
   5-0  &   16    & 49.96 &    -8.50       &0.43 &      \textbf{Chung05}   \cite{Chung05}    \\
   6-0   &  20     & 3.60  &    0.15&       0.49  &     \textbf{Tan17}   \cite{Tan17}      \\
   7-0    & 14 &     6.34   &  -1.55 &      0.46   &    \textbf{Balashov23}  \cite{Balashov23}  \\
   4-1  &   25  &    6.78    & -1.48  &     0.46    &   \textbf{Borkov20a}  \cite{Borkov20a}    \\
   4-1   &   1   &   1.98     & 1.73   &    0.88     &  \textbf{Wojtewicz13} \cite{Wojtewicz13}  \\
   5-1    &  2    & 10.04 &     2.73    &   0.27      & \textbf{Camparque15} \cite{Campargue15}  \\
   5-1  &   18     &10.03  &    1.54     &  0.35       &\textbf{Bordet20} \cite{Bordet20}  \\
   5-4   &   2  &    2.52   &   1.60      & 0.64&       \textbf{Weisbach73} \cite{Weisbach73}   \\
   6-5    &  3   &   1.80    &  2.14       &1.31 &      \textbf{Weisbach73}  \cite{Weisbach73}  \\
   11-10 &   1    &  2.91     & 4.41&       1.52  &     \textbf{Weisbach73}  \cite{Weisbach73}  \\
   \hline
   \multicolumn{6}{l}{$^a$Not used in the fitting.}\\
   \multicolumn{6}{l}{$^b$Possible misprints in the experimental uncertainties of individual lines,} \\
   \multicolumn{6}{l}{\emph{e.g.} 0.03\% for line P11.}
   \end{tabular}
    \label{tab:Fit_quality}
\end{table}

We remind that the description of the experimental data on the transition intensities for the fundamental and low-overtone bands does not impose strict restrictions on the analytical properties of the model functions since any errors in the functions result in similar errors in the calculated intensities. The situation changes cardinally for the high-overtone transitions because of the specific effect of cancellation, \emph{i.e.} the integral for the transition-dipole moment (TDM) can be orders of magnitude smaller than the integrand containing the product of two wave functions and the DMF.

The first necessary condition is that the potential must have the repulsive branch that becomes much larger than the energy difference between the initial and final vibrational states. This is always associated with a singularity, either the Coulomb-repulsion pole at $r=0$ or the one at $r=-\infty$ in the Morse-like potentials. The repulsive branch results in the exponential decay of the intensities with the overtone energy as described by the NIDL with possible anomalies \cite{Medvedev12}. The second  condition is that both the potential and dipole moment must not have any non-physical singularities in a vicinity of the real axis where the integration occurs since they would provide large contribution to the TDM integral \cite{Landau77} deteriorating the NIDL. The third condition is that the dipole moment must be a smooth function of $r$ not only at the real axis but all over the complex plane because the physical dipole-moment operator is just a linear function of the distances between the charges for both the real and complex distance variables.
Here, we apply these considerations for the analysis of the three model DMFs constructed in Sec. \ref{Sec:fit}.

\begin{figure}[htbp]
    \centering
    \includegraphics[scale=0.25]{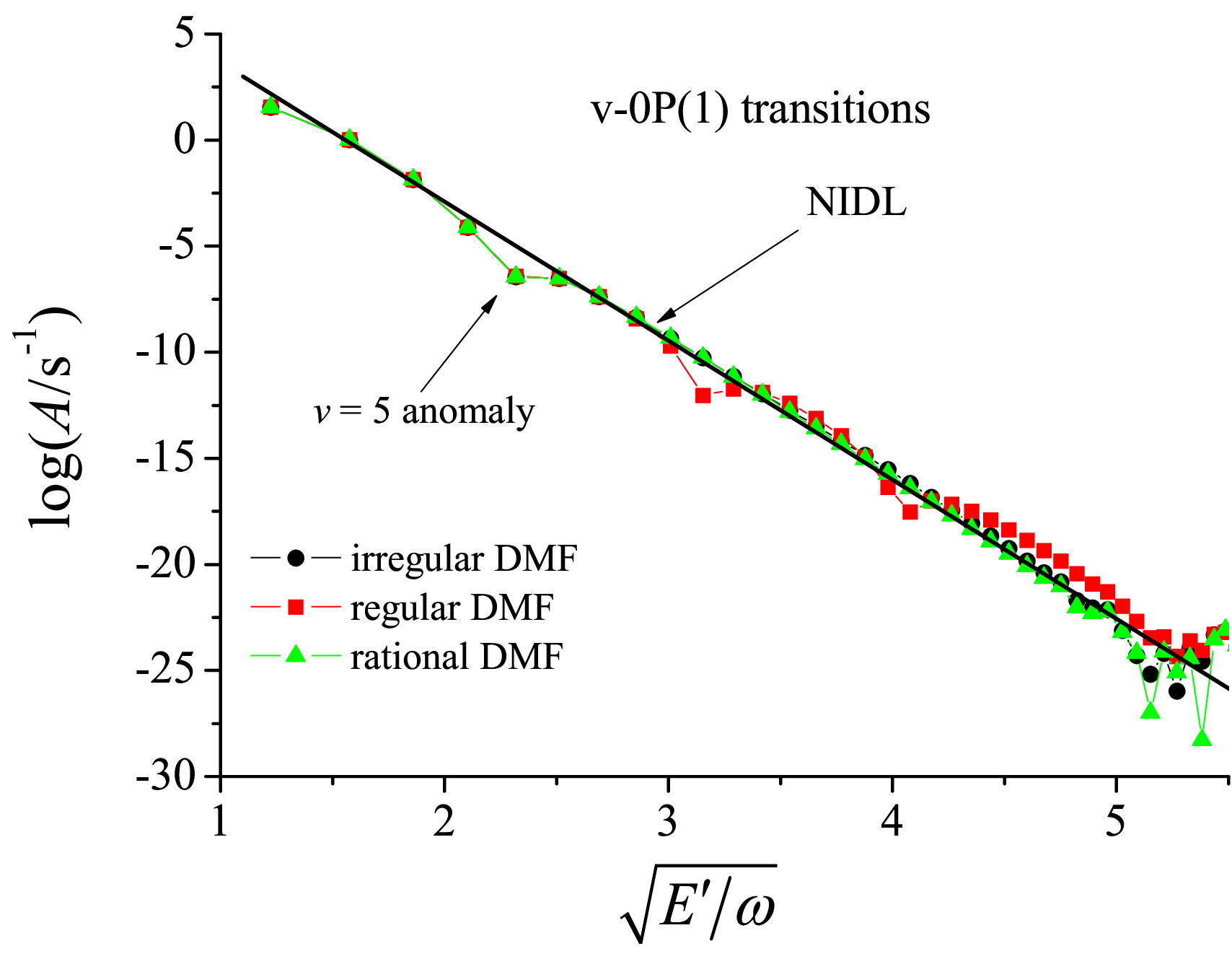}
    \includegraphics[scale=0.25]{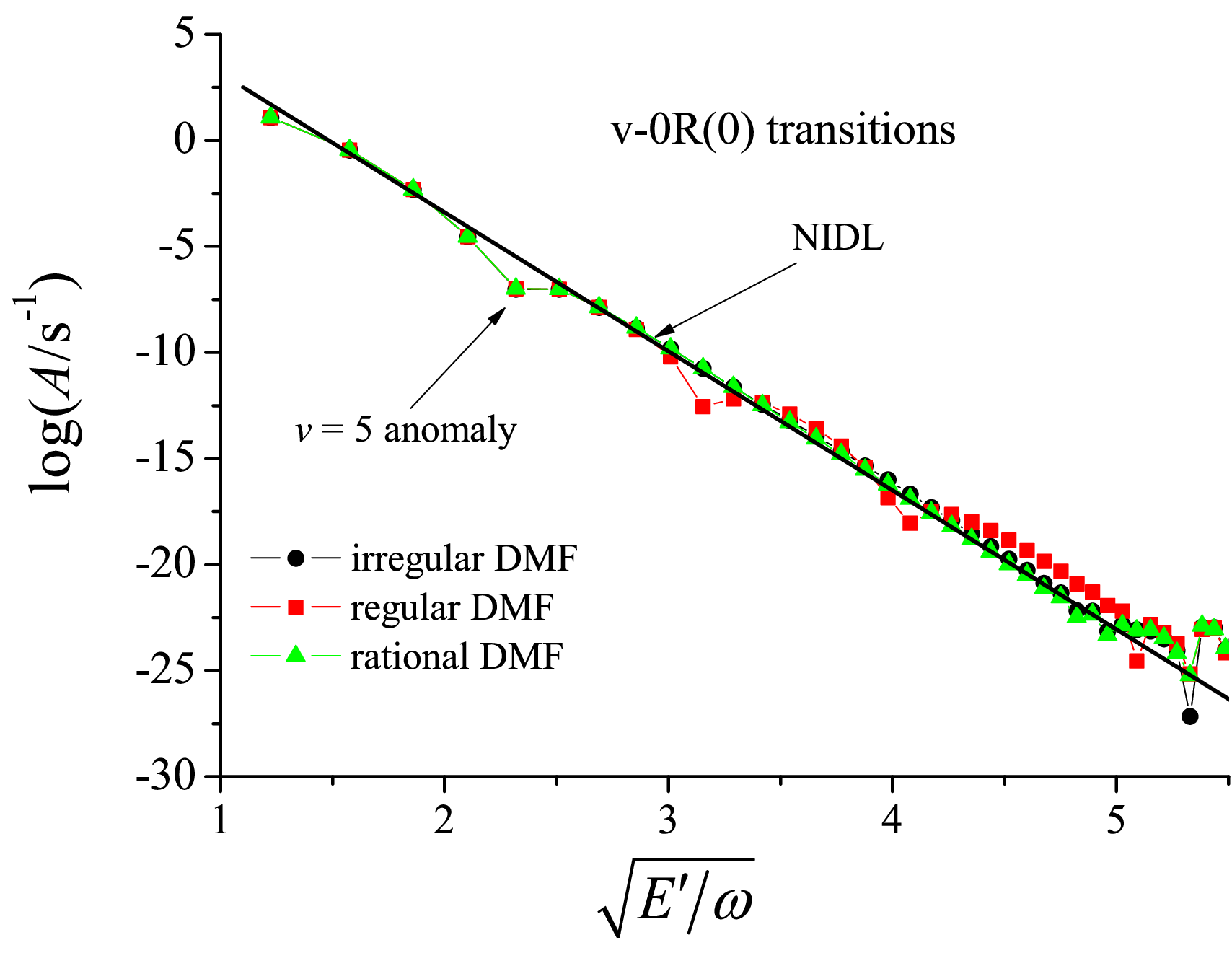}
    \caption{Intensities of the $v$-0 ($v$ = 1,...,40) transitions in the P and R branches in the NIDL coordinates: decimal logarithm of the Einstein-$A$ coefficient \emph{vs} square root of the upper-state energy, $E^\prime$, in units of the vibrational frequency, $\omega=2143$ cm$^{-1}$. The NIDL lines are drawn over the overtone transitions ($v\ge2$) ignoring the anomalies. The $v$ = 5 anomaly was predicted in Ref. \cite{Medvedev85JMS}.}
    \label{fig:NIDL}
\end{figure}

The intensity distributions calculated with three model functions are shown in Fig. \ref{fig:NIDL}. It is seen that the NIDL is valid for the irregular and rational DMFs. For the regular one, the points are shifted up from the NIDL appreciably taking into account that the division value along the ordinate axis is 5.

\begin{figure}[htbp]
    \centering
    \includegraphics[scale=0.25]{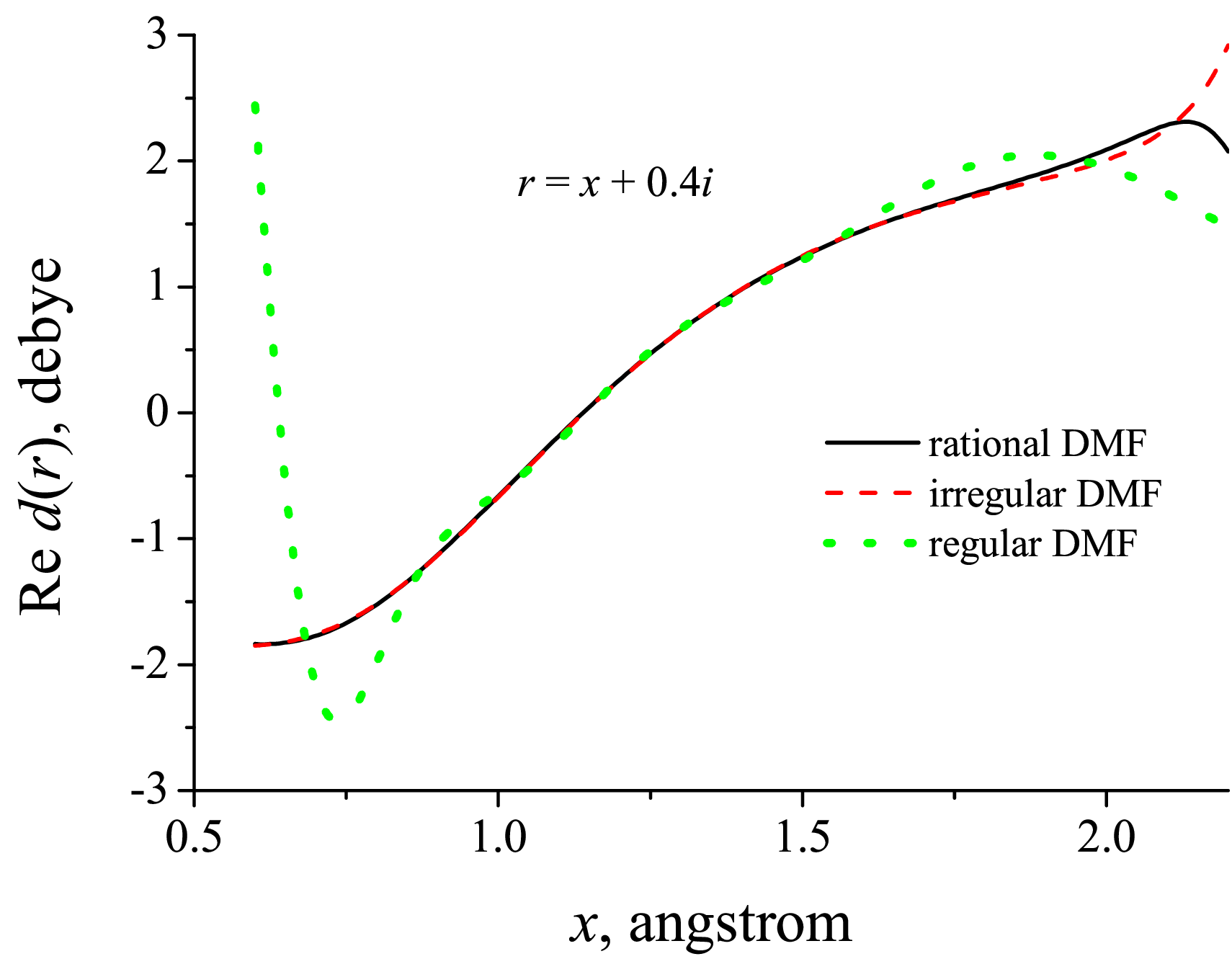}
    \includegraphics[scale=0.25]{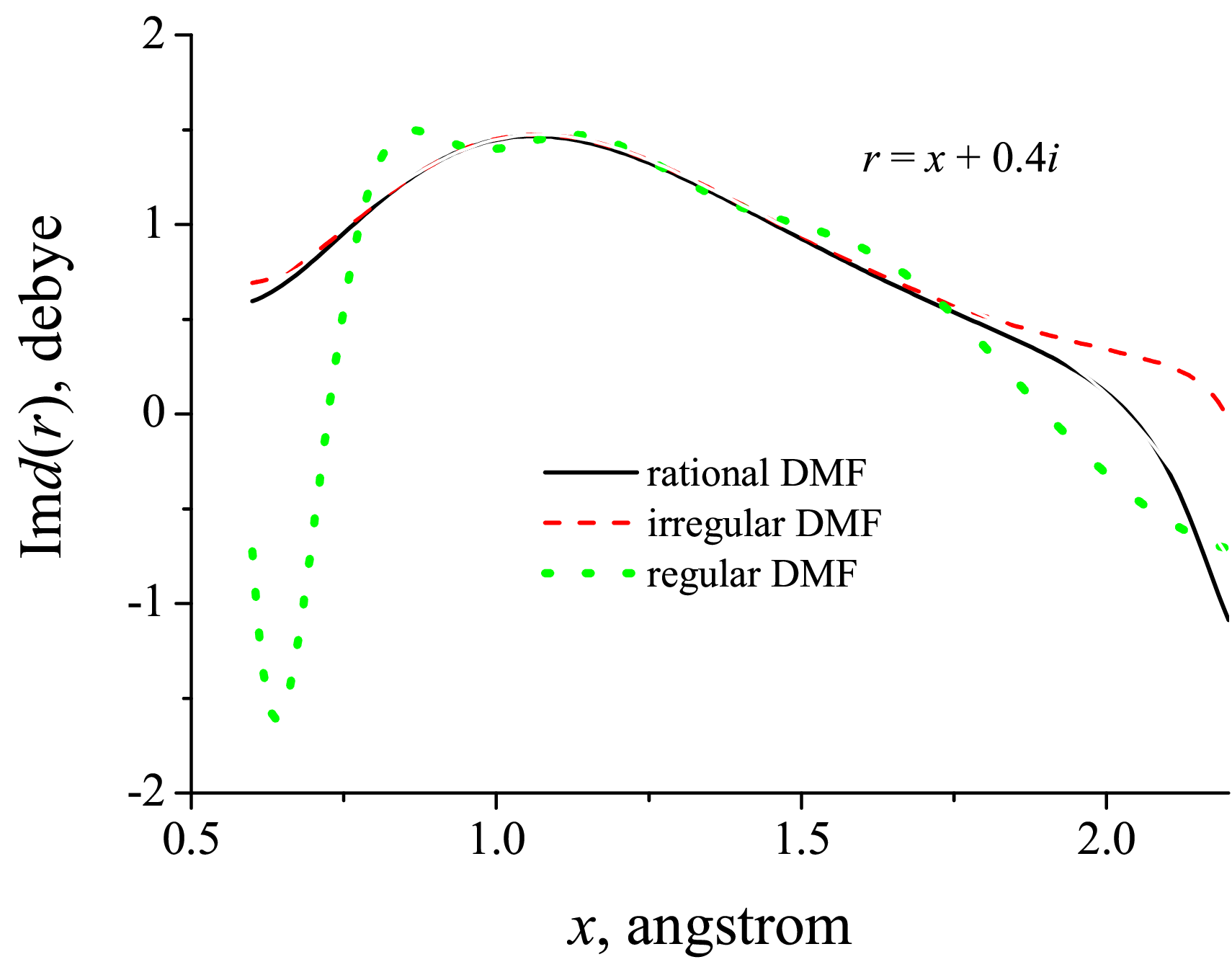}
    \caption{Real and imaginary parts of the DMFs in the complex plane along the line Im $r$ = 0.4 \AA.}
    \label{fig:complex}
\end{figure}

The behavior of the DMFs in the complex plane is shown in Fig. \ref{fig:complex}. The irregular and rational DMFs behave smoothly whereas the regular one shows non-physical behavior that affects the intensities and thereby deteriorates the NIDL. This suggests that the regular DMF is unable to predict the unobserved transitions correctly.

Thus, we conclude that both the irregular and rational DMFs can be used for predictions of the intensities of unobserved transitions. In particular, the 1-0 band can be predicted for comparison with the new high-precision measurements announced in the recent update of the HITRAN database \cite{Gordon25}. The predicted 1-0 band intensities are shown in Table \ref{tab:Band10predictions}, and the comparison with the current and previous HITRAN databases are provided in Supplementary material and in Fig. \ref{fig:B_1-0_comp}. The difference between the HITRAN2020 and HITRAN2016 data is due to the fact that the intensities in the HITRAN2020 were scaled down by 2\% for better agreement with the measurements in \textbf{Devi18}. However, the present calculations show that the latter contain systematic errors. The agreement of the present 1-0 band intensities with HITRAN2016 is about 0.5\%.

\begin{table}[htbp]
    \centering
    \caption{The 1-0 and 8-0 band intensities in units of cm/molecule predicted by the irregular DMF (100\% abundance, 296 K)}
    \vspace{5pt}
    \begin{tabular}{c|c|c|c|c|c}
    $m$ & $S_{1-0}$ & $S_{8-0}$ & $m$ & $S_{1-0}$ & $S_{8-0}$ \\
    \hline
   1  &   9.6582E-20  &  1.970E-31   &   -1  &   9.4419E-20  &  1.971E-31  \\ 
   2  &   1.8996E-19  &  3.830E-31   &   -2  &   1.8155E-19  &  3.833E-31  \\ 
   3  &   2.7503E-19  &  5.481E-31   &   -3  &   2.5696E-19  &  5.486E-31  \\ 
   4  &   3.4739E-19  &  6.842E-31   &   -4  &   3.1731E-19  &  6.852E-31  \\ 
   5  &   4.0375E-19  &  7.859E-31   &   -5  &   3.6053E-19  &  7.873E-31  \\ 
   6  &   4.4214E-19  &  8.504E-31   &   -6  &   3.8598E-19  &  8.523E-31  \\ 
   7  &   4.6202E-19  &  8.780E-31   &   -7  &   3.9431E-19  &  8.804E-31  \\ 
   8  &   4.6418E-19  &  8.716E-31   &   -8  &   3.8730E-19  &  8.743E-31  \\ 
   9  &   4.5058E-19  &  8.358E-31   &   -9  &   3.6754E-19  &  8.389E-31  \\ 
  10  &   4.2398E-19  &  7.769E-31   &  -10  &   3.3811E-19  &  7.802E-31  \\ 
  11  &   3.8766E-19  &  7.016E-31   &  -11  &   3.0224E-19  &  7.051E-31  \\ 
  12  &   3.4503E-19  &  6.167E-31   &  -12  &   2.6300E-19  &  6.203E-31  \\ 
  13  &   2.9931E-19  &  5.283E-31   &  -13  &   2.2306E-19  &  5.318E-31  \\ 
  14  &   2.5335E-19  &  4.416E-31   &  -14  &   1.8460E-19  &  4.449E-31  \\ 
  15  &   2.0942E-19  &  3.604E-31   &  -15  &   1.4918E-19  &  3.634E-31  \\ 
  16  &   1.6915E-19  &  2.873E-31   &  -16  &   1.1781E-19  &  2.901E-31  \\ 
  17  &   1.3359E-19  &  2.240E-31   &  -17  &   9.0971E-20  &  2.264E-31  \\ 
  18  &   1.0320E-19  &  1.708E-31   &  -18  &   6.8711E-20  &  1.728E-31  \\ 
  19  &   7.8010E-20  &  1.274E-31   &  -19  &   5.0787E-20  &  1.291E-31  \\ 
  20  &   5.7725E-20  &  9.301E-32   &  -20  &   3.6745E-20  &  9.441E-32  \\ 
  21  &   4.1824E-20  &  6.649E-32   &  -21  &   2.6033E-20  &  6.760E-32  \\ 
  22  &   2.9679E-20  &  4.654E-32   &  -22  &   1.8063E-20  &  4.740E-32  \\ 
  23  &   2.0631E-20  &  3.191E-32   &  -23  &   1.2278E-20  &  3.256E-32  \\ 
  24  &   1.4052E-20  &  2.143E-32   &  -24  &   8.1777E-21  &  2.192E-32  \\ 
  25  &   9.3791E-21  &  1.410E-32   &  -25  &   5.3375E-21  &  1.446E-32  \\ 
  26  &   6.1357E-21  &  9.096E-33   &  -26  &   3.4146E-21  &  9.345E-33  \\ 
  27  &   3.9346E-21  &  5.749E-33   &  -27  &   2.1413E-21  &  5.922E-33  \\ 
  28  &   2.4736E-21  &  3.562E-33   &  -28  &   1.3165E-21  &  3.679E-33  \\ 
  29  &   1.5247E-21  &  2.163E-33   &  -29  &   7.9364E-22  &  2.241E-33  \\ 
  30  &   9.2162E-22  &  1.288E-33   &  -30  &   4.6916E-22  &  1.339E-33  \\ 
  \hline
    \end{tabular}
    \label{tab:Band10predictions}
\end{table}

\begin{figure}[htbp]
    \centering
    \includegraphics[scale=0.25]{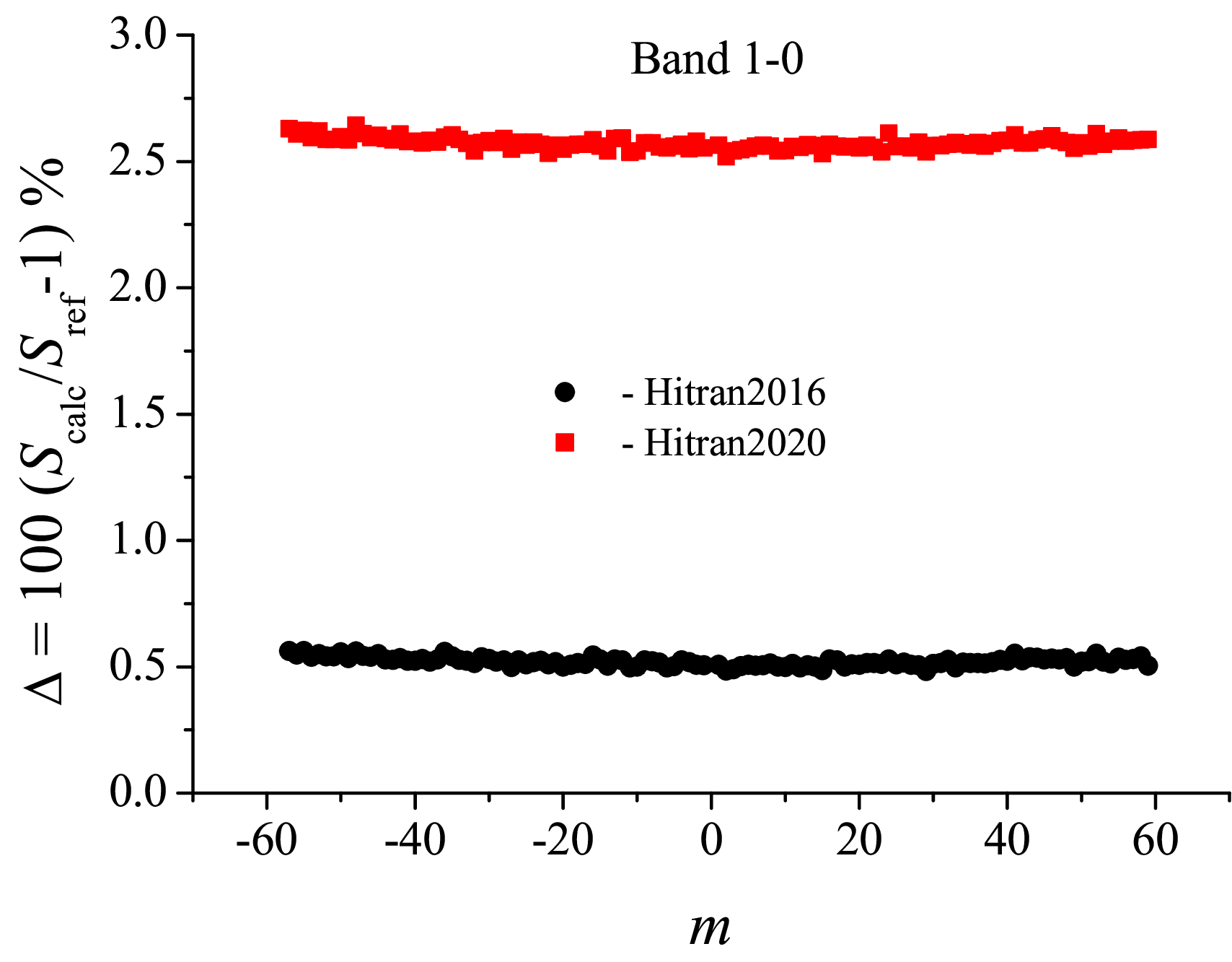}
    \caption{Comparison of the 1-0 band intensities calculated with the irregular DMF (Fit 2) and the HITRAN data.}
    \label{fig:B_1-0_comp}
\end{figure}

\section{Predictions for higher overtones}

\begin{figure}[ht]
    \centering
    \includegraphics[scale=0.25]{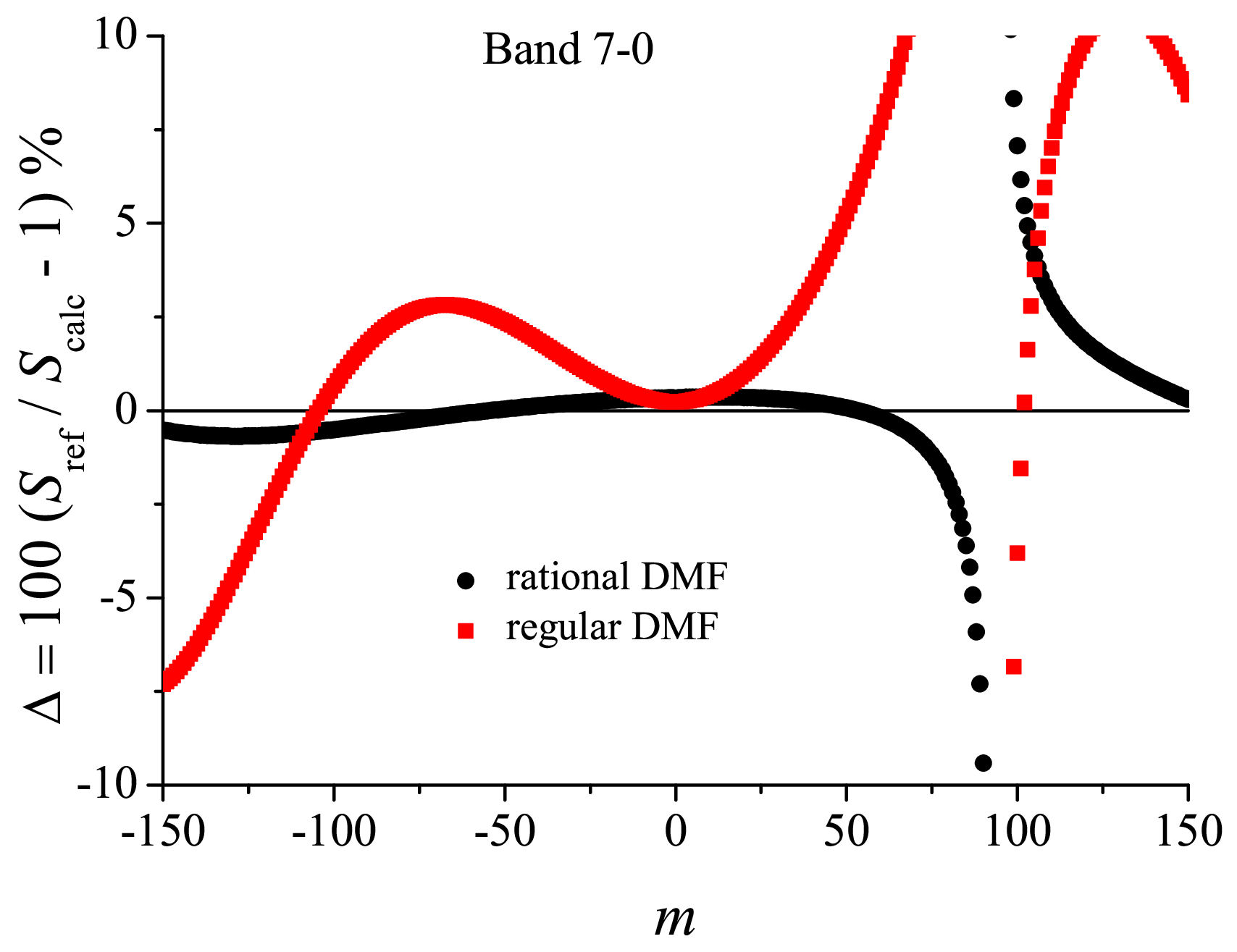}
    \caption{Comparison of the 7-0 band intensities calculated with the rational or regular DMFs, $S_\textrm{ref}$, and the irregular DMF, $S_\textrm{calc}$. The anomaly is at $m=94.$}
    \label{fig:B70predictions}
\end{figure}

\begin{figure}[htbp]
    \centering
    \includegraphics[scale=0.25]{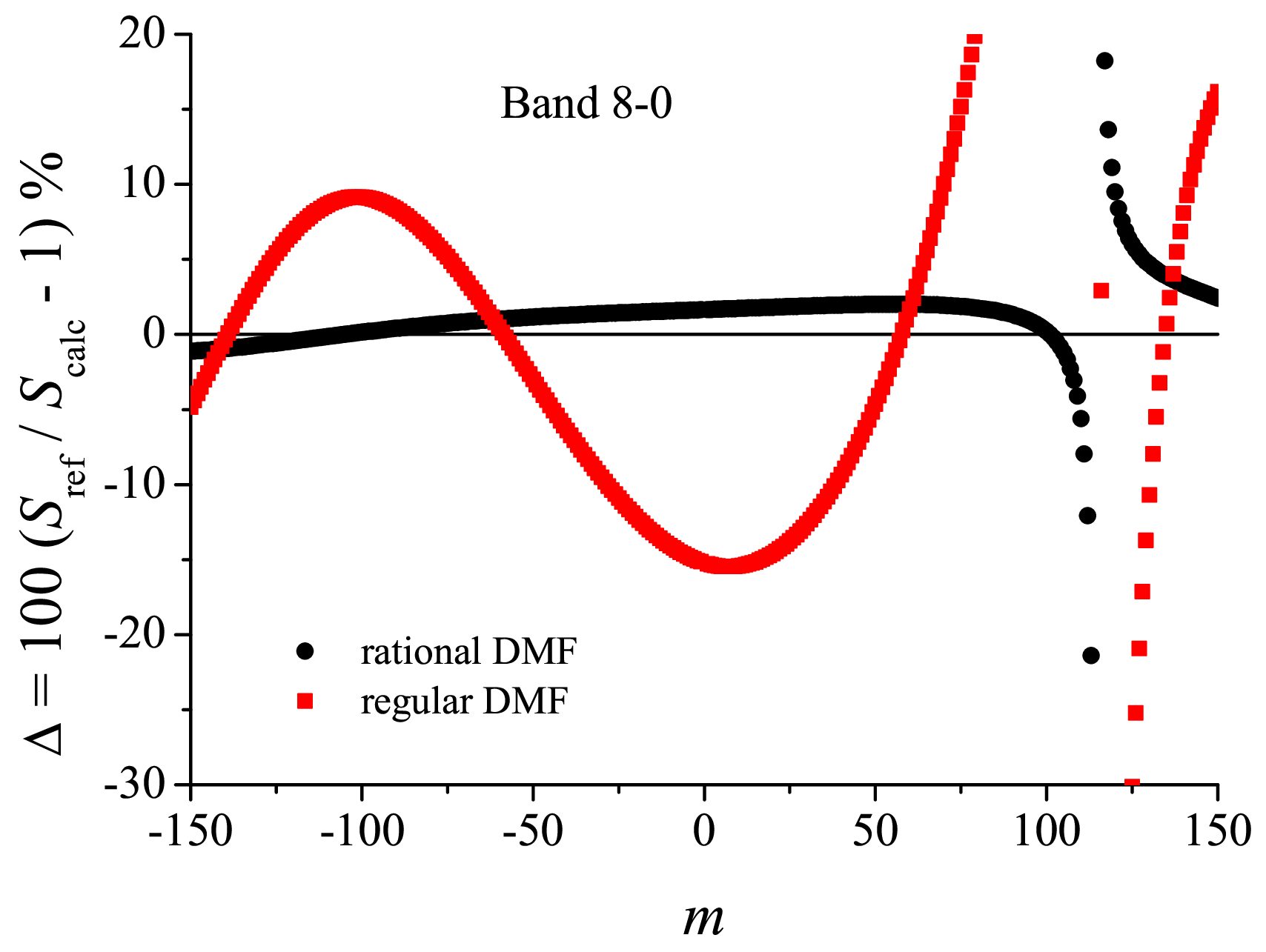}
    \caption{Comparison of the 8-0 band intensities calculated with the rational or regular DMFs, $S_\textrm{ref}$, and the irregular DMF, $S_\textrm{calc}$. The anomalies are at $m=115$-117.}
    \label{fig:Band80comparison}
\end{figure}

\begin{figure}[htbp]
    \centering
    \includegraphics[scale=0.25]{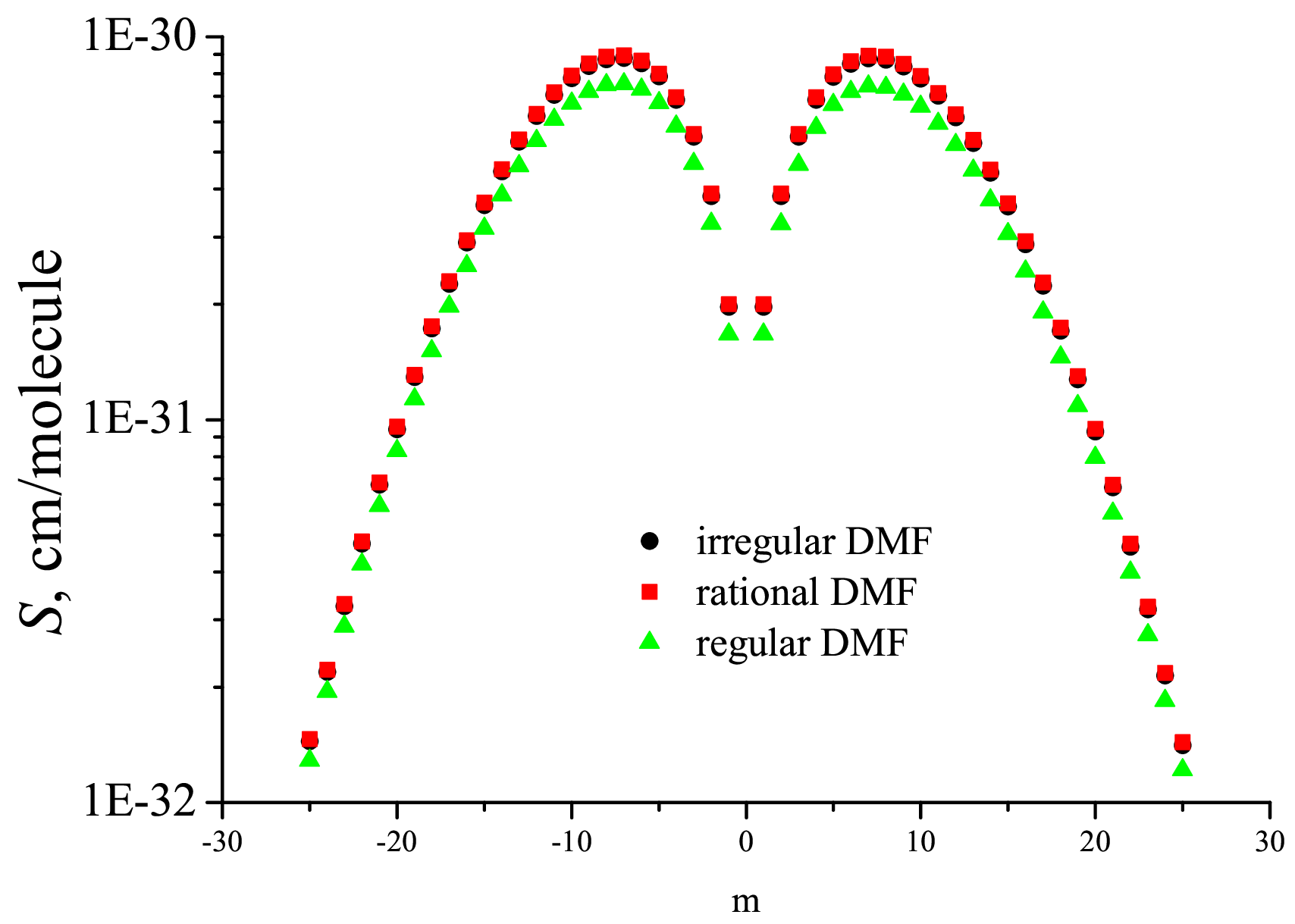}
    \caption{The intensities of the 8-0 band predicted with the three DMFs.}
    \label{fig:Band80S}
\end{figure}

Figure \ref{fig:Band_7-0,Fit2} shows that the 7-0 band intensities calculated with the present three DMFs describe the experiment within the experimental uncertainties. Figure \ref{fig:B70predictions} shows that the irregular and rational DMFs give similar predictions for the unobserved transitions within 1\% at $m\in[-150,70]$, far enough from the anomaly at $m=94$, whereas the regular DMF gives significantly larger differences.

Figures \ref{fig:Band80comparison} and \ref{fig:Band80S} show the relative and absolute predictions for the 8-0 band. Notably, the irregular and rational DMFs predict the intensities differing by less than 2\% in the wide interval of $-150\le m\le100$, whereas the regular DMF results in larger differences, up to 20\%, yet well below the conceivable experimental uncertainty.
When $v$ increases further, the differences between the irregular and rational DMFs predictions are also increasing, but very moderately; \emph{e.g.}, for the 14-0 band, they remain within 40\%.

Table \ref{tab:Band10predictions} and Fig. \ref{fig:Band80S} contain predictions for the next yet unobserved 8-0 band made with the irregular DMF. The difference in the intensities with the rational DMF are within 2\% over the wide interval of $m\in[-150,100]$.

The full line list for $^{12}$C$^{16}$O calculated with the irregular DMF along with the FORTRAN code to calculate the irregular DMF are included in Supplementary material.

\section{Conclusions}

The main conclusion is that the previous data on the 1-0 band \cite{Zou02,Devi18} contradict the new high-precision data \cite{Bielska22,Hodges25,Balashov23}. Therefore, the old data have been excluded from the new fitting of the irregular DMF constructed in Ref. \cite{Medvedev22}. The other important issue is that the uncertainties in the \emph{ab initio} DMF assigned in Ref. \cite{Meshkov22} and used in \cite{Medvedev22} are too low, therefore, they have been increased hugely in the new fitting. As a result, all the experimental and theoretical data have been described well, and predictions for the 1-0 and 8-0 bands have been made. 

Our analysis has shown that the irregular, rational, and regular DMFs constructed in Refs. \cite{Medvedev22,Meshkov22} and refitted here satisfactorily describe the old and new experimental and theoretical data (except for the old data for the 1-0 band), yet the regular DMF \cite{Meshkov22} does not satisfy the additional requirements for valuable predictions. The irregular and rational DMFs do satisfy the requirements equally well, therefore they both can be used for predictions. We select the irregular DMF for the calculations of the line list. 

The new line list calculated for the principal isotopologue of CO presents essential improvement as compared to that of Ref. \cite{Medvedev22} since it describes the new high-precision experimental and theoretical data satisfactorily and therefore is more reliable than the previous one. Moreover, the predictions for the higher overtones made with two dipole-moment functions, the irregular and rational ones having absolutely differing analytical properties, coincide with each other essentially better than they did earlier \cite{Medvedev22} because the new experimental data having unprecedented high accuracy impose severe restrictions on the behavior of the functions fitted to them.

\section*{Acknowledgement}

This work was performed under state task, Ministry of Science and Education, Russian Federation, state registration number 124013000760-0.

\newpage

\bibliography{CO_update_2025,Overtones-Converted-Recovered-2}
\bibliographystyle{elsarticle-num}

\end{document}